\documentclass[12pt,preprint]{aastex}

\usepackage{amssymb,amsmath,epsfig}
\allowdisplaybreaks

\begin{document}

\title{Stability of a Class of Non-Static Axial Self-Gravitating
Systems in $f(R)$ Gravity}

\author{M. Sharif\thanks{msharif.math@pu.edu.pk}  and Z. Yousaf\thanks{zeeshan.math@pu.edu.pk}}
\affil{Department of Mathematics, University of the Punjab,\\
Quaid-e-Azam Campus, Lahore-54590, Pakistan.}

\begin{abstract}
In this paper, we analyze stability regions of a non-static
restricted class of axially symmetric spacetime with anisotropic
matter distribution. We consider $f(R)=R+{\epsilon}R^2$ model and
assume hydrostatic equilibrium of the axial self-gravitating system
at large past time. Considering perturbation from hydrostatic phase,
we develop dynamical as well as collapse equations and explore
dynamical instabilities at Newtonian and post-Newtonian regimes. It
is concluded with the help of stiffness parameter, $\Gamma_1$, that
radial profile of physical parameters like pressure anisotropy,
energy density and higher curvature terms of the $f(R)$ model affect
the instability ranges.
\end{abstract}


\section{Introduction}

Astrophysical latest observations of various experiments like
Supernova (Ia) \citep{t1, t1a}, weak gravitational lensing of
distant galaxies \cite{t2}, Wilkinson Microwave Anisotropy Probe
\citep{t3}, galactic cluster X-rays emission \citep{t4} etc indicate
the accelerated expansion of the universe. This expansion is assumed
to be due to some unusual type of matter referred to dark energy
(DE) whose existence can be inferred from its gravitational effects.
Many models have been presented to explain the mysterious nature of
DE. The cosmic acceleration issue can be better explained using the
modified gravity theories such as $f(R),~f(R,T),~f(G)$ etc. The
$f(R)$ gravity theory \citep{t6, t6a} is the simplest modification
of general relativity (GR) obtained by introducing an arbitrary
function dependence on the Ricci scalar $R$.

There have been theoretical evidences from several physical
processes that show the importance of local anisotropies in pressure
\citep{t8, t8a}. \citet{t10} analyzed the role of slow rotation in
the astronomical anisotropic spherical stars. \citet{t10a} studied
that strong magnetic fields within the compact spheres cause to
generate pressure anisotropy. \citet{t12} examined the effects of
tangential as well as radial components of pressure in the collapse
of relativistic objects within quasi-spherical model.

The analysis on the final collapse phase is of great interest for
many astrophysicists. \citet{t13} explored collapsing process of a
homogenous dust cloud which gives rise to black hole (BH).
\citet{t14} analyzed perfect matter distribution collapse and
concluded that its final phase will be a BH. \citet{t14a} examined
the naked singularity formation with mass parameter in f(R)
background. \citet{t14b} investigated the evolution of collapsing
mechanisms in stellar bodies with $\epsilon R^m$ gravity. Recently,
Sharif and his collaborators \citep{t15a, t15b, t15c, t15d} have
examined the dynamics of collapsing compact bodies and provided
several interesting results through a relationship between
inhomogeneous energy density and Weyl tensor.

The study of gravitational collapse has also been investigated in
modified gravity theories. \citet{t17} worked on BH solution of the
spherical collapsing body in $f(R)$ gravity with a positive constant
Ricci curvature condition. \citet{t18} explored the collapsing
sphere and presented BH solutions with negative constant Ricci
scalar scenario. \citet{t19} studied spherical BH models in the
context of non-constant Ricci scalar condition. \citet{t20}
investigated the collapsing spherical astronomical fluid
distribution in $f(R)$ gravity and concluded that the $f(R)$ gravity
terms delay the black hole formation.

The dynamical instability of astronomical spherical objects like
supermassive stars against radial perturbations and their final
phase has been a subject of keen interest. \citet{t21} presented a
formalism to discuss the dynamical instability of the isotropic
ideal collapsing sphere. \citet{t22} explored the stability of the
spherical dissipative collapsing body and found that heat
dissipation makes the system stable. Many authors \citep{t23a, t23b,
t23c} examined the role of anisotropic pressure, viscosity and
radiation density on the instability of spherical collapse at both
Newtonian (N) as well as post-Newtonian (pN) approximations.
\citet{t24} described the pN cosmological dynamics of the
irrotational dust cloud. \citet{t124} explored modified
Lan\'{e}-Emden equation through N limit of metric $f(R)$ gravity and
analyzed hydrostatic phases of stellar structures.

\citet{t125} investigated Jeans instability for self-gravitating
stellar systems through weak field approximation $f(R)$ gravity
coupled with ideal matter distribution. Sharif and his collaborators
\citep{t25a, t25b, t26, t25c} discussed the instability ranges of
the collapsing stars and found that kinematical quantities like the
higher curvature terms of $f(R)$ gravity, shearfree condition etc
cause the system less unstable thereby slowing down the collapse
rate. \citet{t126} numerically solved modified Lan\'{e}-Emden
equation in order to discuss the stability of some unexplored exotic
stellar structures in $f(R)$ background.

To explore the mysterious phenomenon of dark universe, variety of
$f(R)$ models are extensively studied. \citet{t26a} discussed
various formalisms to discuss expansion of accelerating universe.
\citet{t26b} suggested several generic functions of $f(R,G)$ and
$f(R)$ gravity theories to analyzed dark aspects of Universe during
late-time era. \citet{t26c} presented some viable $f(R)$ gravity
models in both Einstein and Jordan frames. \citet{t127} proposed
exponential-type model which provides the accelerating cosmic
solutions lacking a future singularity.

\citet{t128} introduced some well-consistent $f(R)$ models
describing both inflation as well as late-time acceleration of the
universe. The more interesting $f(R)$ models along with their viable
conditions have been proposed by \citet{t131}. \citet{t32} claimed
that the study of gravitational collapse in the presence of dark
source $f(R)$ terms gives an interesting results about the cosmic
acceleration. Therefore, it would be very worthwile to discuss the
dynamical instability regions of collapsing systems with $f(R)$
models.

\citet{t27} presented several solutions of axially symmetric body
with the help of spherical configuration of matter in $f(R)$
gravity. \citet{t28} found axially symmetric analytic solutions and
concluded that under spherical limit, their solution represent the
Schwarzschild metric. \citet{t130} proved that inflationary and/or
dark energy higher curvature $f(R)$ terms support the
anti-evaporation of Schwarzschild-de Sitter black hole on classical
level. \citet{t129} used cosmological patch technique to discuss
Nariai black holes future evolution with power-law $f(R)$ theory and
concluded that instabilities of such stellar structure depends upon
the specific choice of model.

In this paper, we perform stellar stability analysis for a
non-static axially symmetric restricted spacetime with anisotropic
matter configuration through the perturbation technique in $f(R)$
gravity. The outline of the paper as follows. In the next section,
we develop the field equations as well as dynamical equations for
anisotropic fluid distribution with viable $f(R)$ model. In section
\textbf{3}, we formulate collapse equation using perturbation scheme
while section \textbf{4} establishes the instability ranges in the N
and pN regimes. In the last section, we conclude our results.

\section{Anisotropic Source and Field Equations}

The Einstein-Hilbert action in $f(R)$ gravity can be written as
\begin{equation*}\label{1a}
S_{f(R)}=\frac{1}{2\kappa}\int d^4x\sqrt{-g}f(R)+S_M,
\end{equation*}
where $f(R)$ is a non-linear real function of the curvature $R$,
$\kappa$ is the coupling constant and $S_M$ is the matter action.
The corresponding field equations in metric formalism are
\begin{equation}\label{1b}
f_RR_{\alpha\beta}-\frac{1}{2}fg_{\alpha\beta}-\nabla_{\alpha}
\nabla_{\beta}f_R+g_{\alpha\beta}{\Box}f_R={\kappa}T_{\alpha\beta},
\end{equation}
where $\nabla_\alpha$ indicates covariant derivative and
$\Box=\nabla^\alpha\nabla_{\alpha}$. We can develop a formulation in
the form of Einstein field equations as
\begin{equation}\label{1c}
G_{\alpha\beta}=\frac{\kappa}{f_R}(\overset{(D)}
{T_{\alpha\beta}}+T_{\alpha\beta}),
\end{equation}
where
\begin{equation*}\label{1d}
\overset{(D)}{T_{\alpha\beta}}=\frac{1}{\kappa}\left\{\frac{f-Rf_R}{2}
g_{\alpha\beta}+\nabla_{\alpha}\nabla_{\beta}f_R-\Box
f_Rg_{\alpha\beta}\right\},
\end{equation*}
is the effective stress energy-momentum tensor. The trace of
Eq.(\ref{1b}) yields
\begin{equation*}\label{1e}
Rf_R+3{\Box}f_R-2f={\kappa}T,
\end{equation*}
which under constant curvature condition gives
\begin{equation*}\label{1f}
Rf_R-2f=0.
\end{equation*}
This represents de-Sitter point, a vacuum solution, i.e., $T=0$.

The most general axial symmetric metric in spherical coordinates can
be reduced to the form \citep{t29}
\begin{equation}\label{1}
ds^2=-A^2(t,r,\theta)dt^{2}+B^2(t,r,\theta)(dr^{2}+r^2d\theta^2)
+C^2(t,r,\theta)d\phi^2.
\end{equation}
This is the restricted non-static axial symmetry since we have
ignored the terms describing reflection as well as those
characterizing the rotations about the symmetry axis, i.e.,
$dtd\phi$. The consideration of ``rotation" and ``reflection" along
with four independent metric coefficients in $f(R)$ gravity could
not be handled analytically. We have excluded such terms in our
metric just for the sake of simplicity. It is worthy to mention that
recently, \citet{t28her} have examined dissipative gravitational
collapse of non- static axially symmetric spacetime in GR. They
performed analysis by restricting the axial symmetry to reflection
and neglecting the meridional motions. We consider that the
collapsing axial symmetric body is filled with anisotropic matter
whose energy-momentum tensor is \citep{t28}
\begin{eqnarray}\nonumber
T_{\alpha\beta}&=&(\mu+P)V_\alpha V_\beta-(K_{\alpha}K_\beta
-\frac{1}{3}h_{\alpha\beta})(P_{zz}-P_{xx})+P
g_{\alpha\beta}\\\label{2}
&-&(P_{zz}-P_{yy})(L_{\alpha}L_\beta-\frac{1}{3}h_{\alpha\beta})
+2K_{(\alpha}L_{\beta)}P_{xy},
\end{eqnarray}
where
\begin{equation*}\nonumber
h_{\alpha\beta}=g_{\alpha\beta}+V_\alpha V_\beta, \quad
P=\frac{1}{3}(P_{yy}+P_{xx}+P_{zz}),
\end{equation*}
$\mu$ is the energy density and $P_{xx},~P_{yy},~P_{zz}$ are the
different pressures with $P_{yy}\neq P_{xx}\neq P_{zz}$ and
$P_{xy}=P_{yx}$. Also, $V_\alpha,~K_\alpha$ and $L_\alpha$ are the
four velocity, unit four-vectors, respectively and $\alpha,~ \beta$
are the Lorentz indices. In comoving coordinate system, we have
\begin{equation}\label{3}
V_{\beta}=-A\delta^{0}_{\beta},\quad
L_{\beta}=rB\delta^{2}_{\beta},\quad K_{\beta}=B\delta^{1}_{\beta}.
\end{equation}

The quantities controlling the kinematics of the matter distribution
are the shear tensor $\sigma_{\alpha\beta}$, the four acceleration
$a_{\alpha}$ and the expansion scalar $\Theta$. The non-zero values
of these variables are given as
\begin{eqnarray*}\nonumber
a_{1}&=&\frac{A'}{A},\quad a_{2}=\frac{A^\theta}{A},\quad
\Theta=\frac{1}{A}\left(\frac{\dot{C}}{C}+\frac{2\dot{B}}{B}
\right),\quad\sigma_{11}=\frac{\sigma}{3}B^2,\\\nonumber
\sigma_{22}&=&\frac{\sigma}{3}r^2B^2,\quad
\sigma_{33}=-\frac{2\sigma}{3}C^2,\quad
\sigma=-\frac{1}{A}\left(\frac{\dot{C}}{C}-\frac{\dot{B}}{B}\right),
\end{eqnarray*}
where dot, prime and superscript $\theta$ represent derivatives with
respect to time, radius and $\theta$, respectively. The
corresponding Ricci scalar becomes
\begin{eqnarray}\nonumber
R&=&2\left[\frac{1}{B^2}\left\{\frac{A''}{A}+\frac{A'C'}{AC}+\frac{B''}{B}
+\frac{1}{r}\left(\frac{C'}{C}+\frac{B'}{B}-\frac{A'}{A}\right)
-\frac{B'^2}{B^2}+\frac{C''}{C}\right.\right.\\\nonumber
&+&\left.\frac{1}{r^2}\left(\frac{B^{\theta\theta}}{B}-\frac{B^{\theta2}}{B^2}
+\frac{A^{\theta\theta}}{A}+\frac{C^{\theta\theta}}{C}+\frac{A^\theta
C^\theta}{AC}\right)\right\}\\\label{4}
&+&\left.\frac{1}{A^2}\left\{
\frac{\dot{A}}{A}\left(\frac{2\dot{B}}{B}+\frac{\dot{C}}{C}\right)
-\frac{\dot{B}}{B}\left(\frac{\dot{C}}{C}+\frac{\dot{B}}{B}\right)
-\frac{2\ddot{B}}{B}-\frac{\ddot{C}}{C}\right\}\right],
\end{eqnarray}
The corresponding $f(R)$ field equations take the form
\begin{eqnarray}\nonumber
&&\frac{\dot{B}}{BA^2}\left(\frac{2\dot{C}}{C}
+\frac{\dot{B}}{B}\right)-\frac{1}{B^2}\left[\frac{1}{r}
\left(\frac{C'}{C}+\frac{B'}{B}\right)-\left(\frac{B'}{B}
\right)^2+\frac{B''}{B}+\frac{C''}{C}
+\frac{1}{r^2}\right.\\\nonumber
&&\times\left.\left\{\frac{C^{\theta\theta}}{C}-\left(
\frac{B^\theta}{B}\right)^2+\frac{B^{\theta\theta}}{B}
\right\}\right]=\frac{{\kappa}}{f_R}
\left[\mu+\frac{1}{\kappa}\left\{\frac{f-Rf_R}{2}+\frac{f''_R}{B^2}
-\frac{\dot{f_R}}{A^2}\left(\frac{\dot{2B}}{B}\right.
\right.\right.\\\label{5}
&&\left.\left.\left.+\frac{\dot{C}}{C}\right)
+\frac{f'_R}{B^2}\left(\frac{C'}{C}-\frac{2B'}{B}+\frac{1}{r}\right)
+\frac{f_R^\theta}{B^2r^2}\left(\frac{C^\theta}{C}-\frac{2B^\theta}{B}\right)
+\frac{f_R^{\theta\theta}}{B^2r^2}\right\}\right],
\\\label{6}
&&\frac{A'}{A}\left(\frac{\dot{B}}{B}+\frac{\dot{C}}{C}\right)
+\frac{\dot{B}}{B}\left(\frac{B'}{B}+\frac{C'}{C}\right)
-\frac{\dot{B}'}{B}-\frac{\dot{C}'}{C}=
\frac{\dot{f'_R}}{f_R}-\frac{A'\dot{f_R}}{Af_R}
-\frac{\dot{B}f'_R}{Bf_R}, \\\label{7}
&&\frac{A^\theta}{A}\left(\frac{\dot{B}}{A}+\frac{\dot{C}}{C}\right)
+\frac{\dot{B}}{B}\left(\frac{B^\theta}{B}+\frac{C^\theta}{C}\right)
-\frac{\dot{B}^\theta}{B}-\frac{\dot{C}^\theta}{C}=
\frac{\dot{f_R}^\theta}{f_R}-\frac{A^\theta\dot{f_R}}{Af_R}
-\frac{\dot{B}f^\theta_R}{Bf_R}, \\\nonumber
&&\frac{1}{A^2}\left[\frac{\dot{B}{\dot{A}}}{BA}
-\frac{\ddot{C}}{C}-\frac{\ddot{B}}{B}+\frac{\dot{C}}{C}\left(
\frac{\dot{A}}{A}-\frac{\dot{B}}{C}\right)
\right]+\frac{1}{B^2}\left[\frac{A'B'}{AB}
+\frac{C'}{C}\left(\frac{A'}{A}+\frac{B'}{B}\right)\right.\\\nonumber
&&+\left.\frac{1}{r}\left(\frac{C'}{C}+\frac{A'}{A}\right)
+\frac{1}{r^2}\left\{\frac{A^{\theta\theta}}{A}+\frac{
C^\theta}{C}\left(\frac{A^\theta}{A}-\frac{B^\theta}{B}\right)
+\frac{C^{\theta\theta}}{C}-\frac{A^\theta
B^\theta}{AB}\right\}\right]\\\nonumber
&&=\frac{\kappa}{f_R}\left[P_{xx}
-\frac{1}{\kappa}\left\{\frac{f-Rf_R}{2}-\frac{\ddot{f_R}}{A^2}
+\frac{f_R^{\theta\theta}}{B^2r^2}+\frac{\dot{f_R}}{A^2}
\left(\frac{\dot{A}}{A}-\frac{\dot{B}}{B}+\frac{\dot{C}}{C}\right)
+\frac{f'_R}{B^2}\right.\right.\\\label{8}
&&\left.\left.\times\left(\frac{A'}{A}-\frac{B'}{B}-\frac{1}{r}
+\frac{C'}{C}\right)+\frac{f_R^\theta}{B^2r^2}
\left(\frac{A^\theta}{A}-\frac{3B^\theta}{B}+\frac{C^\theta}{C}\right)\right\}\right]
,\\\nonumber &&\frac{1}{A^2}\left[\frac{\dot{B}{\dot{A}}}{BA}
-\frac{\ddot{C}}{C}-\frac{\ddot{B}}{B}+\frac{\dot{C}}{C}\left(\frac{\dot{A}}{A}
-\frac{\dot{B}}{B}\right)\right]
+\frac{1}{B^2}\left[\frac{C''}{C}+\frac{A''}{A}-\frac{A'B'}{AB}
+\frac{C'}{C}\right.\\\nonumber &&\left.\times\left(\frac{A'}{A}
-\frac{B'}{B}\right)+\frac{1}{r^2}\left\{\frac{B^\theta
A^\theta}{BA}+\frac{C^\theta}{C}\left(\frac{A^\theta
}{A}+\frac{B^\theta}{B}\right)\right\}\right]=\frac{{\kappa}}{f_R}
\left[P_{yy}-\frac{1}{\kappa}\right.\\\nonumber
&&\times\left.\left\{\frac{f-Rf_R}{2}-\frac{\ddot{f_R}}{A^2}
+\frac{f''_R}{B^2}+\frac{\dot{f_R}}{A^2}\left(\frac{\dot{A}}{A}-\frac{\dot{B}}{B}
-\frac{\dot{C}}{C}\right)+\frac{f'_R}{B^2}\left(\frac{A'}{A}-\frac{B'}{B}
+\frac{C'}{C}\right)\right.\right.\\\label{9}
&&\left.\left.+\frac{f_R^\theta}{B^2r^2}\left(\frac{A^\theta}{A}
-\frac{B^\theta}{B}+\frac{C^\theta}{C}\right)\right\}\right],\\\nonumber
&&\frac{1}{A^2}\left[\frac{\dot{B}}{B}\left(\frac{\dot{A}}{A}
+\frac{\dot{B}}{B}\right)+\frac{2\ddot{B}}{B}
\right]+\frac{1}{B^2}\left[\frac{B''}{B}+\frac{A''}{A}
-\left(\frac{B'}{B}\right)^2+\frac{1}{r}\left(\frac{B'}{B}
+\frac{A'}{A}\right)\right. \\\nonumber
&&+\left.\frac{1}{r^2}\left\{\frac{B^{\theta\theta}}{B}+\frac{A^{\theta\theta}}{A}
-\frac{B^{\theta2}}{B^2}\right\}\right]
=\frac{{\kappa}}{f_R}\left[P_{zz}-\frac{1}{\kappa}\left\{
\frac{f-Rf_R}{2}-\frac{\ddot{f_R}}{A^2}+\frac{f''_R}{B^2}
+\frac{f_R^{\theta\theta}}{B^2r^2}\right.\right.\\\label{10}
&&\left.\left.
+\frac{\dot{f_R}}{A^2}\left(\frac{\dot{A}}{A}-\frac{2\dot{B}}{B}\right)
+\frac{f'_R}{B^2}\left(\frac{A'}{A}-\frac{2B'}{B}-\frac{1}{r}\right)
+\frac{f_R^{\theta\theta}}{B^2r^2}\left(\frac{A^\theta}{A}
-\frac{2B^\theta}{B}\right)\right\}\right],\\\nonumber
&&\frac{1}{B^2}\left[\frac{1}{r}\left\{-\frac{C^{'\theta}}{C}-\frac{A^{'\theta}}{A}
+\frac{B'}{B}\left(\frac{A^\theta}{A}+\frac{C^\theta}{C}\right)
+\frac{B^\theta}{B}\left(\frac{C'}{C}+\frac{A'}{A}\right)
\right\}\right.\\\label{11}
&&+\left.\left(\frac{C^\theta}{C}+\frac{A^\theta}{A}
\right)\frac{1}{r^2}\right]=\frac{\kappa}{f_R}\left[P_{xy}+\frac{1}{\kappa}
\left(f^{'\theta}_R-\frac{B^{\theta}f'_R}{B}-\frac{f^\theta_R}{r}
-\frac{B'f^\theta_R}{B}\right)\right].
\end{eqnarray}

The dynamical equations describe how parameters of the
self-gravitating collapsing objects evolve with time and radius. The
corresponding dynamical equations can be found through
\begin{eqnarray*}
\left(T^{\alpha\beta}+\overset{(D)}{T^{\alpha\beta}}\right)_{;\beta}V_\alpha=0,\quad
\left(T^{\alpha\beta}+\overset{(D)}{T^{\alpha\beta}}\right)_{;\beta}L_\alpha=0,\quad
\left(T^{\alpha\beta}+\overset{(D)}{T^{\alpha\beta}}\right)_{;\beta}K_\alpha=0,
\end{eqnarray*}
which lead to
\begin{eqnarray}\label{12}
&&\frac{\dot{\mu}}{A}+(\mu+P_{zz})\frac{\dot{C}}{AC}
+\frac{\dot{B}}{AB}(2\mu+
P_{yy}+P_{xx})+\frac{D_0(t,r,\theta)}{{\kappa}A}=0,\\\nonumber
&&\frac{P_{xx}'}{B}+\frac{A'}{AB}(P_{xx}+\mu)-\frac{B'}{B^2}(P_{yy}-P_{xx})
-\frac{C'}{BC}(P_{zz}-P_{xx})+\frac{1}{rB}\\\label{13}
&&\times\left\{P_{xy}\left(\frac{C^\theta}{C}+
\frac{2B^\theta}{B}+\frac{A^\theta}{A}\right)
+P^\theta_{xy}-P_{yy}+P_{xx}\right\}+\frac{D_1(t,r,\theta)}{{\kappa}B}=0,\\\nonumber
&&\frac{rP^\theta_{yy}}{B}+\frac{rA^\theta}{AB}(P_{yy}+\mu)
-\frac{rB^\theta}{B^2}(P_{xx}-P_{yy})-\frac{rC^\theta}{BC}(P_{zz}-P_{yy})
+\frac{2rP_{xy}}{B}\\\label{14}
&&+\left[P_{xy}'+P_{xy}\left(\frac{C'}{C}+\frac{2B'}{B}+\frac{A'}{A}\right)
\right]\frac{r^2}{B}+\frac{rD_2(t,r,\theta)}{{\kappa}B}=0,
\end{eqnarray}
where $D_0,~D_1$ and $D_2$ are given in Appendix \textbf{A}. Many
inflation models in the early universe are established on scalar
fields coming from supergravity and superstring theories. The first
model of inflation was suggested by Starobinsky which corresponds to
the conformal anomaly in quantum gravity \citep{t27} given by
\begin{equation}\label{15}
f(R)=R+{\epsilon}R^{2}.
\end{equation}
This model was suggested both as a model for dark matter \citep{t32}
as well as an inflationary candidate \citep{t31} which can lead to
the accelerated universe expansion due to $R^2$ term. It is
mentioned here that $\epsilon=\frac{1}{6M^2}$ which is proposed for
dark matter model. It is worth mentioning here that the value of $M$
is figured out as $2.7\times10^{-12}GeV$ with
$\epsilon\leq2.3\times10^{22}Ge/V^2$ \citep{t30}. General relativity
is recovered for ${\epsilon}=0$ that corresponds to classically
stable BH \citep{t33}. In $f(R)$ gravity, all these characteristics
are also observed thus the stability condition for this theory takes
the form $[{\epsilon}(1+2{{\epsilon}}R)]^{-1}\geq0$. Recently,
slow-roll inflation has been discussed through $f(R)$ polynomial
models \citep{tpoly, tpoly1}.

\section{Perturbation Scheme}

Here, we use the perturbation technique \citep{t22} to perturb the
dynamical equations and Ricci scalar upto first order in $\alpha$,
where $0<\alpha\ll1$. The system is assumed to be entirely in the
state of hydrostatic equilibrium, but upon evolution, time
dependence factor $T(t)$ appears in all the functions that are
controlling the kinematics of the system. We further consider that
all the metric coefficients possess the same time dependence that
imparts similar time dependence on the Ricci scalar. Notice that
$T(t)$ is not a trace of the energy-momentum tensor instead an
arbitrary function of time. The fluid and metric variables are
perturbed as
\begin{eqnarray}\label{16}
A(t,r,\theta)&=&A_0(r,\theta)+{\alpha}T(t)a(r,\theta),\\\label{17}
B(t,r,\theta)&=&B_0(r,\theta)+{\alpha}T(t)b(r,\theta),\\\label{18}
C(t,r,\theta)&=&C_0(r,\theta)+{\alpha}T(t)c(r,\theta),\\\label{19}
\mu(t,r,\theta)&=&\mu_0(r,\theta)+{\alpha}\bar{\mu}(t,r,\theta),\\\label{20}
P_{xy}(t,r,\theta)&=&P_{xy0}(r,\theta)+{\alpha}\bar{P}_{xy}(t,r,\theta),\\\label{21}
P_{xx}(t,r,\theta)&=&P_{xx0}(r,\theta)+{\alpha}\bar{P}_{xx}(t,r,\theta),\\\label{22}
P_{yy}(t,r,\theta)&=&P_{yy0}(r,\theta)+{\alpha}\bar{P}_{yy}(t,r,\theta),\\\label{23}
P_{zz}(t,r,\theta)&=&P_{zz0}(r,\theta)+{\alpha}\bar{P}_{zz}(t,r,\theta),\\\label{24}
R(t,r,\theta)&=&R_0(r,\theta)+{\alpha}T(t)e(r,\theta),\\\label{25}
f(t,r,\theta)&=&R_0(1+{\epsilon}R_0)+{\alpha}T(t)e(r,\theta)
(1+2{\epsilon}R_0),\\\label{26}
f_R(t,r,\theta)&=&(1+2{\epsilon}R_0)+2{\alpha}{\epsilon}T(t)e(r,\theta),
\end{eqnarray}
where $R_0$ is the static part of Eq.(\ref{4}) and is computed as
\begin{eqnarray}\nonumber
&&R_0=\frac{2}{B^2_0}\left\{\frac{C''_0}{C_0}
+\frac{B''_0}{B_0}+\frac{A'_0C'_0}{A_0C_0} +\frac{A''_0}{A_0}
+\frac{1}{r}\left(-\frac{A'_0}{A_0}+\frac{C'_0}{C_0}+\frac{B'_0}{B_0}\right)
-\frac{B'^2_0}{B^2_0}\right.\\\label{27} &&+\left.\frac{1}{r^2}
\left(\frac{C^{\theta\theta}_0}{C_0}+\frac{B^{\theta\theta}_0}{B_0}
+\frac{A^{\theta\theta}_{0}}{A_0}+\frac{C^\theta_0A^\theta_0}{A_0C_0}
-\frac{B^{\theta2}_0}{B^2_0}\right)\right\}.\\\nonumber
\end{eqnarray}

The static part of the field equations (\ref{5})-(\ref{11}) are
obtained by using Eqs.(\ref{16})-(\ref{26}) as follows
\begin{eqnarray}\nonumber
&&-\frac{1}{B^2_0}\left[\frac{B''_0}{B_0}-\left(\frac{B'_0}{B_0}\right)^2
+\frac{1}{r}\left(\frac{B'_0}{B_0}
+\frac{C'_0}{C_0}\right)+\frac{C''_0}{C_0}
\frac{1}{r^2}\left\{\frac{B^{\theta\theta}_0}{B_0}
-\left(\frac{B^{\theta}_0}{B_0}\right)^2\right.\right.\\\nonumber
&&+\left.\left.
\frac{C^{\theta\theta}_0}{C_0}\right\}\right]=\frac{\kappa}
{1+2{\epsilon}R_0}\left[\mu_0+\frac{2{\epsilon}}{\kappa}
\left\{-\frac{R_0^2}{4}+\frac{R''_0}{B_0^2}+\frac{R'_0}{B_0^2}
\left(\frac{C'_0}{C_0}-\frac{2B'_0}{B_0}\right.\right.\right.\\\label{28}
&&\left.\left.\left.+\frac{1}{r}\right)
+\frac{R_0^\theta}{B_0^2r^2}\left(\frac{C_0^\theta}{C_0}-\frac{2B_0^\theta}
{B_0}\right)+\frac{R_0^{\theta\theta}}{B_0^2r^2}\right\}\right],
\\\nonumber
&&\frac{1}{B^2_0}\left[\frac{B'_0A'_0}{A_0B_0}+\frac{C'_0}{C_0}
\left(\frac{A'_0}{A_0}+\frac{B'_0}{B_0}\right)
+\frac{1}{r}\left(\frac{C'_0}{C_0}+\frac{A'_0}{A_0}\right)
+\frac{1}{r^2}\left\{\frac{C^{\theta\theta}_0}{C_0}+\frac{A^{\theta\theta}_0}{A_0}
+\frac{C^{\theta}_0}{C_0}\right. \right.\\\nonumber
&&\times\left.\left.\left(\frac{A^{\theta}_0}{A_0}
-\frac{B^{\theta}_0}{B_0}\right) -\frac{A^{\theta}_0
B^{\theta}_0}{A_0B_0}\right\}\right]
=\frac{\kappa}{1+2{\epsilon}R_0}\left[P_{xx0}-\frac{2{\epsilon}}{\kappa}
\left\{-\frac{R_0^2}{4}+\frac{R_0^{\theta\theta}}{B_0^2r^2}
\right.\right.\\\label{29}
&&\left.\left.+\frac{R'_0}{B_0^2}\left(\frac{A'_0}{A_0}-\frac{B'_0}{B_0}-\frac{1}{r}
+\frac{C'_0}{C_0}\right)+\frac{R_0^\theta}{B_0^2r^2}
\left(\frac{A_0^\theta}{A_0}-\frac{3B_0^\theta}
{B_0}+\frac{C_0^\theta}{C_0}\right)\right\}\right],\\\nonumber
&&\frac{1}{B^2_0}\left[\frac{C''_0}{C_0}+\frac{A''_0}{A_0}
+\frac{C'_0}{C_0}\left(\frac{A'_0}{A_0}
-\frac{B'_0}{B_0}\right)-\frac{A'_0B'_0}{A_0B_0}
+\frac{1}{r^2}\left\{\frac{B^{\theta}_0
A^{\theta}_0}{A_0B_0}+\frac{C^{\theta}_0}{C_0}
\left(\frac{A^{\theta}_0}{A_0}\right.\right.\right.
\\\nonumber
&&\left.\left.\left.+\frac{B^{\theta}_0}{B_0}\right)\right\}\right]
=\frac{{\kappa}}{1+2{\epsilon}R_0}\left[P_{yy0}+\frac{2{\epsilon}}{\kappa}
\left\{\frac{R_0^2}{4}-\frac{R''_0}{B_0^2}-\frac{R'_0}{B_0^2}
\left(\frac{A'_0}{A_0}-\frac{B'_0}{B_0}
+\frac{C'_0}{C_0}\right)\right.\right.\\\label{30}
&&\left.\left.-\frac{R_0^\theta}{B_0^2r^2}
\left(\frac{A_0^\theta}{A_0}-\frac{B_0^\theta}
{B_0}+\frac{C_0^\theta}{C_0}\right)\right\}\right],\\\nonumber
&&\frac{1}{B^2_0}\left[\frac{B''_0}{B_0}+\frac{A''_0}{A_0}
-\left(\frac{B'_0}{B_0}\right)^2+\frac{1}{r}\left(\frac{B'_0}{B_0}
+\frac{A'_0}{A_0}\right)+\frac{1}{r^2}\left\{\frac{B^{\theta\theta}_0}{B_0}+
\frac{A^{\theta\theta}_0}{A_0}
-\frac{B^{\theta2}_0}{B_0^2}\right\}\right]\\\nonumber
&&=\frac{\kappa}{1+2{\epsilon}R_0}\left[P_{zz0}-\frac{2{\epsilon}}{\kappa}
\left\{-\frac{R_0^2}{4}+\frac{R^{\theta\theta}_0}{B_0^2}
+\frac{R''_0}{B_0^2}+\frac{R'_0}{B_0^2}
\left(\frac{A'_0}{A_0}-\frac{2B'_0}{B_0}
-\frac{1}{r}\right)\right.\right.\\\label{31}
&&\left.\left.+\frac{R_0^\theta}{B_0^2r^2}
\left(\frac{A_0^\theta}{A_0}-\frac{2B_0^\theta}{B_0}\right)\right\}\right]
,\\\nonumber
&&\frac{1}{B^2_0}\left[\frac{1}{r}\left\{-\frac{A^{'\theta}_0}{A_0}
+\frac{B'_0C^{\theta}_0}{B_0C_0}-\frac{C^{'\theta}_0}{C_0}
+\frac{B^{\theta}_0}{B_0}\left(\frac{A'_0}{A_0}+\frac{C'_0}{C_0}\right)
+\frac{A^{\theta}_0
B'_0}{A_0B_0}\right\}+\frac{1}{r^2}\left(\frac{A^{\theta}_0}{A_0}
\right.\right.\\\label{32}
&&\left.\left.+\frac{C^{\theta}_0}{C_0}\right)\right]
=\frac{\kappa}{1+2{\epsilon}R_0}\left[P_{xy0}+\frac{2{\epsilon}}{\kappa}
\left\{R^{'\theta}_0-\frac{B^{\theta}_0}{B_0}R'_0
-\frac{R^\theta_0}{r}-\frac{B'_0}{B_0}R_0^\theta\right\}\right] .
\end{eqnarray}
The dynamical equations (\ref{12})-(\ref{14}) under static
background leads to
\begin{eqnarray}\nonumber
&&P_{xx0}'+(P_{xx0}+\mu_0)\frac{A'_0}{A_0}-(P_{yy0}-P_{xx0})\frac{B'_0}{B_0}
-(P_{zz0}-P_{xx0})\frac{C'_0}{C_0}+\frac{P_{xy0}}{r}\\\label{33}
&&\times\left(\frac{C^\theta_0}{C_0}
+\frac{2B^\theta_0}{B_0}+\frac{A^\theta_0}{A_0}\right)
+\frac{1}{r}\left(P^\theta_{xy0}-P_{yy0}+P_{xx0}\right)
+\frac{D_{1S}(r,\theta)}{\kappa}=0,\\\nonumber
&&P^\theta_{yy0}+(P_{yy0}+\mu_0)\frac{A^\theta_0}{A_0}
-(P_{xx0}-P_{yy0})\frac{B^\theta_0}{B_0}
-(P_{zz0}-P_{yy0})\frac{C^\theta_0}{C_0}\\\label{34}
&&+2P_{xy0}+r\left\{P_{xy0}\left(\frac{C'_0}{C_0}+
\frac{2B'_0}{B_0}+\frac{A'_0}{A_0}\right)
+P_{xy0}'\right\}+\frac{D_{2S}(r,\theta)}{\kappa}=0,
\end{eqnarray}
where $D_{1S}$ and $D_{2S}$ are static parts of the above equations
given in Appendix \textbf{A}.

The perturbed form of Eqs.(\ref{12})-(\ref{14}) will be
\begin{eqnarray}\label{35}
&&\dot{\bar{\mu}}+\left\{\frac{c}{C_0}(P_{zz0}+\mu_0)
+(2\mu_0+P_{xx0}+P_{yy0})\frac{b}{B_0}+\frac{D_3(r,\theta)}{\kappa}
\right\}\dot{T}=0,\\\nonumber
&&\bar{P}_{xx}'-\left(\frac{a}{A_0}\right)'(P_{xx0}+\mu_0)T
+(\bar{\mu}+\bar{P}_{xx})\frac{A_0'}{A_0}
-T\left(\frac{c}{C_0}\right)'(P_{zz0}-P_{xx0})
\\\nonumber &&-(\bar{P}_{zz}-\bar{P}_{xx})\frac{C_0'}{C_0}
-T\left(\frac{b}{B_0}\right)'(P_{yy0}-P_{xx0})
-\frac{B_0'}{B_0}(\bar{P}_{yy}-\bar{P}_{xx})+\frac{1}{r}
(\bar{P}^\theta_{xy}\\\nonumber
&&-\bar{P}_{yy}+\bar{P}_{xx})+T\left(\frac{c}{C_0}+\frac{2b}{B_0}
+\frac{a}{A_0}\right)^{\theta}\frac{P_{xy0}}{r}
+\frac{\bar{P}_{xy}}{r}\left(\frac{C^\theta_0}{C_0}
+\frac{2B^\theta_0}{B_0}+\frac{A^\theta_0}{A_0}\right)\\\label{36}
&&+\frac{P_1(t,r,\theta)}{\kappa}=0,\\\nonumber
&&\bar{P}^\theta_{yy}+(\bar{P}_{yy}+\bar{\mu})\frac{A^\theta_0}{A_0}
-T\left(\frac{a}{A_0}\right)^{\theta}(P_{yy0}+\mu_0)
-(\bar{P}_{xx}-\bar{P}_{yy})\frac{B^\theta_0}{B_0}\\\nonumber
&&+T\left(\frac{b}{B_0}\right)^{\theta}(P_{xx0}-P_{yy0})
-(\bar{P}_{zz}-\bar{P}_{yy})\frac{C^\theta_0}{C_0}+T
\left(\frac{c}{C_0}\right)^{\theta}(P_{zz0} -P_{yy0})\\\nonumber
&&+rTP_{xy0}\left(\frac{c}{C_0}+\frac{2b}{B_0}
+\frac{a}{A_0}\right)'+r\left(\frac{C_0'}{C_0}+\frac{2}{r}
+\frac{2B_0'}{B_0}+\frac{A_0'}{A_0}\right)\bar{P}_{xy}\\\label{37}
&&+r\bar{P}_{xy}+\frac{D_5(t,r,\theta)}{\kappa}=0.
\end{eqnarray}
The perturbed part of the Ricci scalar becomes
\begin{eqnarray}\nonumber
&&Te=\frac{2T}{B_0^2} \left[\frac{C_0'A_0'}{C_0A_0}\left(
\frac{a'}{A_0'}-\frac{a}{A_0}+\frac{c'}{C_0'}-\frac{c}{C_0}\right)
+\left(\frac{a''}{A_0''}-\frac{a}{A_0}\right)\frac{A_0''}
{A_0}+\left(\frac{b''}{B_0''}-\frac{b}{B_0}\right)\right.\\\nonumber
&&\times\frac{B_0''}{B_0}
-\frac{1}{r}\left(\frac{a}{A_0}-\frac{c}{C_0}-\frac{b}{B_0}\right)'
+\left(\frac{\bar{c}''}{C_0''}-\frac{\bar{c}}{C_0}\right)\frac{C_0''}
{C_0}-2\left(\frac{b}{B_0}\right)'\frac{B_0'}{B_0}+\left(\frac{b}{B_0}
\right)^{\theta}\frac{B^\theta_0}{B_0}\\\nonumber
&&\times\frac{2}{r^2}+\left(\frac{c^{\theta\theta}}
{C^{\theta\theta}_0}-\frac{c}{C_0}\right)\frac{C^{\theta\theta}_0}
{C_0}+\left(\frac{b^{\theta\theta}}{B^{\theta\theta}_0}-\frac{b}{B_0}
\right)\frac{B^{\theta\theta}_0}{B_0}+\left(\frac{a^{\theta\theta}}
{A^{\theta\theta}_0}-\frac{a}{A_0}
\right)\frac{A^{\theta\theta}_0}{A_0}+\frac{C^{\theta}_0A^{\theta}_0}
{C_0A_0}\\\label{38}
&&\left.\times\left(\frac{a^\theta}{A^{\theta}_0}-\frac{a}{A_0}
+\frac{c^\theta}{C^{\theta}_0}-\frac{c}{C_0}\right)\right]
+\frac{2\ddot{T}}{A_0^2}\left(\frac{c}{C_0}-\frac{b}{B_0}
\right)-2Tb\frac{R_0}{B_0},
\end{eqnarray}
which can be written as
\begin{eqnarray}\label{39}
\ddot{T}(t)-{\delta^2(r,\theta)}T(t)=0,
\end{eqnarray}
where $\delta^2$ is given in Appendix \textbf{A}. The solutions of
Eq.(\ref{39}) corresponds to the unstable as well as stable matter
distributions. Since we are aiming to find the unstable range of
collapsing body, so we assume $T(-\infty)=0$. For this purpose, we
have
\begin{equation}\label{40}
T(t)=-\exp({\delta}t),
\end{equation}
where $\delta^2>0$.

We can develop a relation between perturbed pressure components
$\bar{P}_{i}$ and energy density $\bar{\mu}$ through the equation of
state as \citep{t34}
\begin{equation}\label{41}
\bar{P}_{i}=\Gamma_1\frac{P_{i0}}{\mu_0+P_{i0}}\bar{\mu},
\end{equation}
where $\Gamma_1$ is the adiabatic index which is taken as a constant
identity throughout the paper. Using Eqs.(\ref{35}) and (\ref{41}),
it follows that
\begin{eqnarray*}\nonumber
&&\bar{P}_{xx}=\frac{-\Gamma_1}{\mu_0+P_{xx0}}
\left[\frac{c}{C_0}(P_{zz0}+\mu_0)+
(P_{xx0}+2\mu_0+P_{yy0})\frac{b}{B_0}+\frac{D_3}{\kappa}\right]P_{xx0}T,\\\nonumber
&&\bar{P}_{xy}=\frac{-\Gamma_1}{\mu_0+P_{xy0}}
\left[\frac{c}{C_0}(P_{zz0}+\mu_0)+
(P_{xx0}+2\mu_0+P_{yy0})\frac{b}{B_0}+\frac{D_3}{\kappa}\right]P_{xy0}T,\\\nonumber
&&\bar{P}_{yy}=\frac{-\Gamma_1}{\mu_0+P_{yy0}}
\left[\frac{c}{C_0}(P_{zz0}+\mu_0)+
(P_{xx0}+2\mu_0+P_{yy0})\frac{b}{B_0}+\frac{D_3}{\kappa}\right]P_{yy0}T,\\\nonumber
&&\bar{P}_{zz}=\frac{-\Gamma_1}{\mu_0+P_{zz0}}
\left[\frac{c}{C_0}(P_{zz0}+\mu_0)+
(P_{xx0}+2\mu_0+P_{yy0})\frac{b}{B_0}+\frac{D_3}{\kappa}\right]P_{zz0}T.
\end{eqnarray*}
Using these equations in Eq.(\ref{36}), we obtain the collapse
equation as follows
\begin{eqnarray}\nonumber
&&-T\Gamma_1\left[\left\{(P_{zz0}+\mu_0)\frac{c}{C_0}
+(P_{yy0}+P_{xx0})\frac{b}{B_0}+\frac{D_3}{\kappa}\right\}
\left(\frac{P_{xx0}}{\mu_0+P_{xx0}}\right)\right]'\\\nonumber
&&+T\left(\frac{c}{C_0}\right)'(P_{xx0}-P_{zz0})
+T\left\{\Gamma_1\frac{P_{xx0}}{P_{xx0}+\mu_0}+1\right\}\frac{A_0'}{A_0}
\left[-\frac{c}{C_0}(P_{zz0}+\mu_0)\right.\\\nonumber
&&\left.-(P_{yy0}+P_{xx0}+2\mu_0)\frac{b}{B_0}-\frac{D_3}{\kappa}\right]
-T\left(\frac{a}{A_0}\right)'(\mu_0+P_{xx0})
+T\left(\frac{b}{B_0}\right)'\\\nonumber &&
\times\left(P_{xx0}-P_{yy0}\right)
+\Gamma_1T\frac{C_0'}{C_0}\left[\frac{c}{C_0}
\left\{P_{zz0}-\frac{(P_{zz0}+\mu_0)P_{xx0}}
{(P_{zz0}+\mu_0)}\right\}+(P_{xx0}+2\mu_0\right.\\\nonumber
&&\left.+P_{yy0})\left\{\frac{P_{zz0}}{\mu_0+P_{zz0}}
-\frac{P_{xx0}}{P_{xx0}+\mu_0}\right\}\frac{b}{B_0}
+\left\{\frac{P_{zz0}}{\mu_0+P_{zz0}}
-\frac{P_{xx0}}{P_{xx0}+\mu_0}\right\}\right.\\\nonumber
&&\left.\times\frac{D_3}{\kappa}\right]+\left(\frac{1}{r}
+\frac{B'_0}{B_0}\right)\Gamma_1T
\left(\frac{P_{yy0}}{P_{yy0}+\mu_0}
-\frac{P_{xx0}}{P_{xx0}+\mu_0}\right)
\left[\frac{c}{C_0}(P_{zz0}+\mu_0)\right.\\\nonumber
&&\left.+(P_{xx0}+P_{yy0}+2\mu_0)
\frac{b}{B_0}+\frac{D_3}{\kappa}\right]
-\Gamma_1\frac{T}{r}\left[\left\{
\frac{c}{C_0}(P_{zz0}+\mu_0)+(P_{xx0}\right.\right.\\\nonumber
&&\left.\left.+P_{yy0}+2\mu_0) \frac{b}{B_0}+\frac{D_3}{\kappa}
\right\}\frac{P_{xy0}}{(P_{xy0}+\mu_0)}\right]^{\theta}
-\Gamma_1\frac{T}{r}\left(\frac{A_0^\theta}{A_0}
+\frac{2B_0^\theta}{B_0}+\frac{C_0^\theta}{C_0}\right)\\\nonumber
&&\left[\left\{\frac{c}{C_0}(P_{zz0}+\mu_0)+(P_{xx0}+P_{yy0}+2\mu_0)
\frac{b}{B_0}+\frac{D_3}{\kappa}
\right\}\frac{P_{xy0}}{(P_{xy0}+\mu_0)}\right]\\\label{42}
&&+P_{xy0}\frac{T}{r}
\left(\frac{c}{C_0}+\frac{b}{B_0}+\frac{a}{A_0}\right)^\theta
+\frac{T(t)D_4(r,\theta)}{\kappa}=0,
\end{eqnarray}
where $P_1=T(t)D_4(r,\theta)$ ((obtained by using Eq.(\ref{40})).
This equation has fundamental importance in stellar instability
analysis of the collapsing astronomical axial body in $f(R)$
gravity.

\section{Instability Regions}

In this section, we investigate the instability regions of a system
at both N and pN approximations in $f(R)$ gravity. We also find the
importance of adiabatic index $\Gamma_1$ in this framework.

\subsection{Newtonian Approximation}

Here, we consider the N limits as $A_0=1,~B_0=1,$ for the
investigation of instability ranges in the N era. Moreover, we
assume that $P_{xx0}<0$ indicates collapsing fluid and $C_0=r$ as
the radial coordinate. Under these assumptions, the collapse
equation (\ref{42}) turns out to be
\begin{eqnarray}\nonumber
&&-\Gamma_1T\left[\left(\frac{c}{r}+2b\right)P_{xx0}\right]'
+\Gamma_1\frac{T}{r}\left[\left(\frac{c}{r}+2b\right)(P_{yy0}-P_{xx0})\right]
+\Gamma_1T\left[\left(\frac{c}{r}+2b\right)\right.\\\nonumber
&&\times\left.(P_{zz0}-P_{xx0})\right]
-\frac{\Gamma_1T}{r}\left[\left(\frac{c}{r}+2b\right)P_{xy0}\right]^\theta
=a'\mu_0+T\left(\frac{c}{r}\right)'(P_{zz0}-P_{xx0})\\\label{43}
&&+Tb'(P_{yy0}-P_{xx0})
-\frac{T}{r}\left(a+b+\frac{c}{r}\right)^\theta{P_{xy0}}-\frac{D_{4(N_0)}}{\kappa},
\end{eqnarray}
where $D_{4(N_0)}$ represents terms of $D_4$ under N limit with
$C_0=r$. The required dynamical instability range for the collapsing
axial symmetric body is given using Eq.(\ref{40}) as
\begin{equation}\label{44}
\Gamma_1<\frac{a'\mu_0+(P_{zz0}-P_{xx0})\left(\frac{c}{r}\right)'-\frac{D_{4(N_0)}}{\kappa}+\Omega_1(r)
}{\frac{1}{r}\left(\frac{c}{r}+2b\right)
(P_{yy0}-2P_{xx0}+P_{zz0})+\Omega_2(r) },
\end{equation}
where we defined
\begin{align}\nonumber
&\Omega_1=b'(P_{yy0}-P_{xx0})-\frac{P_{xy0}}{r}\left(a+b+\frac{c}{r}\right)^\theta
\\\nonumber
&\Omega_2=-\frac{1}{r}\left[\left(\frac{c}{r}+2b\right)P_{xx0}\right]^{\theta}
-\left[\left(\frac{c}{r}+2b\right)P_{xx0}\right]'.
\end{align}
This suggests that $\Gamma_1$ plays a key role in the investigation
of stability at N approximation. Moreover, the fluid distribution
will be unstable unless (\ref{44}) is satisfied. Under the constant
Ricci scalar condition, i.e., $R=\tilde{R}$ and
$e=\tilde{e}=\textmd{constant}$, we have
\begin{equation}\label{45}
\Gamma_1<\frac{a'\mu_0+(P_{zz0}-P_{xx0})\left(\frac{c}{r}\right)'
+\left(2\delta^2_{(N)}-\tilde{R}\right)\frac{{\epsilon}\tilde{e}}{\kappa}
+\Omega_1}{\frac{1}{r}\left(\frac{c}{r}+2b\right)
(P_{yy0}-2P_{xx0}+P_{zz0})+\Omega_2}.
\end{equation}
For $\epsilon\rightarrow0$, we get the same inequality as above lest
$\zeta=0$ which corresponds to GR solution \citep{t26}.

\subsection{Post-Newtonian Approximation}

For the dynamical range of instability in the pN limit, we assume
\begin{eqnarray}\label{46}
A_0=1-\frac{m_0}{r},\quad B_0=1+\frac{m_0}{r},
\end{eqnarray}
and take the effects upto $O(\frac{m_0}{r})$. Using these quantities
in the collapse equation (\ref{42}), we get the instability range as
\begin{align}\label{47}
\Gamma_1<\frac{\left(1+\frac{m_0}{r}\right)\left(1-\frac{m_0}{r}\right)'
\zeta_1+\Omega_3-\frac{D_{4(pN_0)}}{\kappa}}
{-\frac{\zeta_1}{r}\left[\left(1-\frac{m_0}{r}\right)^\theta
\left(1+\frac{m_0}{r}\right)\frac{P_{xy0}}{P_{xy0}+\mu_0}\right]
+\Omega_4},
\end{align}
where
\begin{eqnarray}\nonumber
&&\zeta_1=\frac{c}{r}(P_{zz0}+\mu_0)+(P_{xx0}+P_{yy0}+2\mu_0)
b\left(1-\frac{m_0}{r}\right)+\frac{D_{3pN_0}}{\kappa},\\\nonumber
&&\Omega_1=\left(b-\frac{bm_0}{r}\right)'(P_{yy0}-P_{xx0})
-\frac{P_{xy0}}{r}\left[a+b+\frac{c}{r}+\frac{m_0}{r}
(a+b)\right]^\theta\\\nonumber
&&+\left(a-\frac{am_0}{r}\right)'(P_{xx0}+\mu_0),\\\nonumber
&&\Omega_2=-\left(1+\frac{m_0}{r}\right)\left(1-\frac{m_0}{r}\right)'
\frac{P_{xx0}\zeta_1}{P_{xx0}+\mu_0}+\left\{\frac{1}{r}
+\left(1+\frac{m_0}{r}\right)'\left(1-\frac{m_0}{r}\right)
\right\}\\\nonumber
&&\times\left(\frac{{\zeta_1}P_{yy0}}{P_{yy0}+\mu_0}
-\frac{{\zeta_1}P_{xx0}}{P_{xx0}+\mu_0}\right)
+\frac{c}{r^2}\left\{P_{zz0}-\frac{(P_{zz0}+\mu_0)
P_{xx0}}{P_{xx0}+\mu_0}\right\}
+\frac{1}{r}\left[\left(b\right.\right.\\\nonumber
&&\left.\left.-\frac{bm_0}{r}\right)(2\mu_0+P_{xx0}
+P_{yy0})+\frac{D_3}{\kappa}\right]\left(\frac{P_{zz0}}
{P_{zz0}+\mu_0}-\frac{P_{xx0}}{P_{xx0}+\mu_0}\right)\\\nonumber
&&-\left(\frac{P_{xx0\zeta_1}}{P_{xx0}+\mu_0}\right)_{,1}
-\left(\frac{P_{xy0\zeta_1}}{P_{xy0}+\mu_0}\right)_{,0}-\frac{\zeta_1}{r}
\left[2\left(1-\frac{m_0}{r}\right)
\left(1+\frac{m_0}{r}\right)^\theta\frac{P_{xy0}}{P_{xy0}+\mu_0}\right],
\end{eqnarray}
where $D_{3pN_0}$ and $D_{4pN_0}$ correspond to those quantities of
$D_{3}$ and $D_{4}$ that are computed under pN limits (mentioned in
Eq.(\ref{46})) with $C_0=r$, respectively. Under constant curvature
condition, we obtain the same inequality as (\ref{47}) with the
difference that $D_{3pN_0}$ and $D_{4pN_0}$ reduces to $\zeta_1$ and
$\zeta_2$, respectively and are given as
\begin{eqnarray*}\nonumber
\zeta_1&=&-\epsilon\tilde{e}\tilde{R}
-2\epsilon\tilde{e}\left(1-\frac{m_0}{r}\right)'\left(1-\frac{m_0}{r}
\right) \left[3\left(1-\frac{m_0}{r}\right)'\left(1+\frac{m_0}{r}
\right)\right.\\\nonumber
&-&2\left(1+\frac{m_0}{r}\right)\left(1-\frac{m_0}{r}\right)
\left.+\frac{2}{r}\right]-\frac{2\epsilon\tilde{e}}{r^2}
\left(1-\frac{m_0}{r}\right)^\theta\left(1-\frac{m_0}{r}
\right)\\\nonumber&\times&\left[3\left(1-\frac{m_0}{r}\right)^\theta
\left(1+\frac{m_0}{r}\right)-2\left(1+\frac{m_0}{r}\right)^\theta
\left(1-\frac{m_0}{r}\right)\right],\\\nonumber
\zeta_2&=&\frac{\epsilon\tilde{e}}{\kappa}\left[2\delta^2\left(1
+\frac{m_0}{r}\right)+2\delta^2\left(1-\frac{m_0}{r}\right)'
\left(1+\frac{3m_0}{r}\right)\right].
\end{eqnarray*}
For $\epsilon\rightarrow0$, we obtain the instability constraint
given in  (\ref{47}) with  $\zeta_1=\zeta_2=0$ This indicates that
our results exactly coincide with GR solution \citep{t26}.

\section{Conclusions}

In this work, we have explored factors which affect the
gravitational collapse of non-static axial matter distribution in
$f(R)$ gravity. Our analysis provides corrections to the usual GR
field equations thus modify the dynamical evolutionary phases of
collapse process. The field equations and the dynamical equations
are formulated that are perturbed through the perturbation scheme.
The collapse equation is then constructed from the dynamical
equations to discuss the instability range by assuming the
relationship of perturbed pressure and energy density through
equation of state.

It is well-known that instability issue of stellar bodies are
explored by following two techniques. First scheme is based on
numerical methods which assists us to interpret realistic collapsing
scenarios. Nevertheless, these results are generally restricted and
depends upon model under consideration. The second scheme yields
analytical solutions, which are relatively easy to examine and
provides useful results in the theory of structure formation of
stellar systems. In this paper, we have used analytical approach to
examine the instability regions at N and pN eras.

We have found the instability ranges for N and pN regimes. For N
regime, the system will be unstable if it satisfies the inequality
(\ref{44}) while for pN regime, it will remain unstable if the
inequality (\ref{47}) is satisfied. The violation of these
inequalities will lead to stable configuration of the axial
symmetry. These constraints indicate that stability of the
collapsing anisotropic axial astronomical matter is controlled by
the radial profile of pressure anisotropy, energy density and the
dark source $f(R)$ terms. The relations (\ref{44}) and (\ref{47})
are quoted in terms of the adiabatic index $\Gamma_1$ that indicate
its importance and compatibility with \citet{t21}. We see that the
adiabatic index depends upon the physical parameters of the fluid
distribution. It is worth mentioning here that anisotropic pressure
disturbs the stability of the axial symmetry and makes the system
more unstable as the evolution proceeds.

It is well-known that the stability of self-gravitating systems has
a direct correspondence with hydrostatic equilibrium conditions.
Equations that describe such equilibrium phases are related to
$f(R)$ field equations which couple the fluid distribution with its
gravitational field. The dark source terms that originate due to
$f(R)$ model (Eq.(\ref{15})) in the field equations decrease the
range of instability thus lags the BH formation. It is seen from
relations (\ref{44}) and (\ref{47}) that extra-order $f(R)$
corrections, i.e., $\epsilon$ terms extend the stability ranges of
stellar structures which is in agreement with \citet{t126}.
Moreover, it is also well-established that the inflationary
candidate $f(R)=R+\epsilon R^2$ terms cause anti-evaporation of the
Schwarzschild-de Sitter black hole (so called Nariai black hole) on
classical level. Finally, all our results lead to GR solutions
\citep{t26} under the limit $\epsilon\rightarrow0$.

\vspace{0.3cm}

\renewcommand{\theequation}{A\arabic{equation}}
\setcounter{equation}{0}
\section*{Appendix A}

The higher curvature terms for Eqs.(\ref{12})-(\ref{14}) are given
as
\begin{eqnarray}\nonumber
D_{0}&=&\frac{1}{A}\left[\frac{f-Rf_R}{2}+\frac{f''_R}{B^2}-\frac{\dot{f_R}}{A^2}
\left(\frac{\dot{2B}}{B}+\frac{\dot{C}}{C}\right)
+\frac{f'_R}{B^2}\left(\frac{C'}{C}
-\frac{2B'}{B}+\frac{1}{r}\right)\right.\\\nonumber
&+&\left.\frac{f^\theta_R}{B^2r^2}\left(\frac{C^\theta}{C}-2\frac{B^\theta}{B}
\right)+\frac{f^{\theta\theta}_R}{B^2r^2}\right]_{,0}-\frac{1}{AB^2}
\left(\dot{f'_R}
-\frac{A'}{A}\dot{f_R}-\frac{\dot{B}}{B}f'_R\right)\\\nonumber
&\times&\left(3\frac{A'}{A}
+2\frac{B'}{B}+\frac{C'}{C}+\frac{1}{r}\right)+\frac{1}{AB^2r^2}\left(
\frac{3A^\theta}{A}+\frac{2B^\theta}{B}+\frac{C^\theta}{C}\right)\\\nonumber
&\times&\left(\dot{f^\theta_R}
-\frac{A^\theta}{A}\dot{f_R}-\frac{\dot{B}}{B}f^\theta_R\right)
+A\left[\frac{(-1)}{A^2B^2}\left(\dot{f'_R}-\frac{A'}{A}\dot{f_R}
-\frac{\dot{B}}{B}f'_R\right)\right]_{,1}\\\nonumber
&+&A\left[\frac{-1}{A^2B^2r^2}
\left(\dot{f^\theta_R}-\frac{A^\theta}{A}\dot{f_R}
-\frac{\dot{B}}{B}f^\theta_R\right)\right]_{,2}+\frac{\dot{C}}{AC}
\left[\frac{f'_R}{B^2}\left(\frac{C'}{C}+\frac{2}{r}-\frac{A'}{A}\right)
\right.\\\nonumber
&-&\left.\frac{f^{\theta}_R}{B^2r^2}\left(\frac{C^\theta}{C}
-\frac{A^\theta}{A}\right)-\frac{\dot{f_R}}{A^2}\left(\frac{\dot{A}}{A}
+\frac{\dot{C}}{C}\right)\right]+\frac{\dot{B}}{AB}
\left[\frac{f''_R}{B^2}+\frac{f^{\theta\theta}_R}{B^2r^2}
+\frac{2\ddot{f_R}}{A^2}\right.\\\label{A1}
&-&\left.\frac{f_R^\theta}{B^2r^2}\left(\frac{A^\theta}{A}\right)
-\frac{2\dot{f_R}}{A^2}\left(\frac{\dot{A}}{A}+\frac{\dot{B}}{B}
+\frac{\dot{C}}{C}\right)-\frac{f'_R}{B^2}\left(\frac{2A'}{A}+\frac{2B'}{B}
+\frac{3}{r}\right)\right], \\\nonumber
D_{1}&=&\frac{(-1)}{B}\left[\frac{f-Rf_R}{2}-\frac{\ddot{f_R}}{A^2}
+\frac{f^{\theta\theta}_R}{B^2r^2}+\frac{f'_R}{B^2}\left(\frac{A'}{A}
-\frac{B'}{B}-\frac{1}{r}+\frac{C'}{C}\right)\right.\\\nonumber
&+&\left.\frac{\dot{f_R}}{A^2}
\left(\frac{\dot{A}}{A}-\frac{\dot{B}}{B}+\frac{\dot{C}}{C}\right)
+\frac{f^\theta_R}{B^2r^2}\left(\frac{A^\theta}{A}-3\frac{B^\theta}{B}
+\frac{C^\theta}{C}\right)\right]_{,1}-\frac{1}{A^2B}\\\nonumber
&\times&\left(\dot{f'_R}
-\frac{A'}{A}\dot{f_R}-\frac{\dot{B}}{B}f'_R\right)\left(\frac{\dot{A}}{A}
+4\frac{\dot{B}}{B}+\frac{\dot{C}}{C}\right)+\frac{1}{B^3r^2}\left(
\frac{A^\theta}{A}+\frac{4B^\theta}{B}\right)\\\nonumber
&\times&\left(f'^\theta_R
-\frac{B^\theta}{B}f'_R-\frac{f^\theta_R}{r}-\frac{B'}{B}f^\theta_R\right)
+B\left[\frac{(-1)}{A^2B^2}\left(\dot{f'_R}-\frac{A'}{A}\dot{f_R}
-\frac{\dot{B}}{B}f'_R\right)\right]_{,0}\\\nonumber
&+&B\left[\frac{1}{B^4r^2}
\left(f'^\theta_R-\frac{B^\theta}{B}f'_R-\frac{f^\theta_R}{r}
-\frac{B'}{B}f^\theta_R\right)\right]_{,2}+\frac{A'}{AB}\left[\frac{f''_R}
{B^2}+\frac{\ddot{f_R}}{A^2}\right.\\\nonumber
&-&\left.\frac{f^{\theta\theta}_R}{B^2r^2}
-\frac{\dot{f_R}}{A^2}\left(\frac{\dot{B}}{B}+\frac{\dot{A}}{A}
+2\frac{\dot{C}}{C}\right)+\frac{f^\theta_R}{B^2r^2}\left(\frac{B^\theta}
{B}-\frac{A^\theta}{A}\right)+\frac{f'_R}{B^2}\left(\frac{2}{r}\right.\right.\\\nonumber
&-&\left.\left.\frac{B'}{B}
-\frac{A'}{A}\right)\right]+\left(\frac{1}{r}+\frac{B'}{B}\right)\frac{1}{B}
\left[\frac{f''_R}
{B^2}-\frac{f^{\theta\theta}_R}{B^2r^2}-\frac{2\dot{f_R}\dot{C}}{A^2C}
-\frac{f'_R}{B^2r}\right.\\\nonumber
&+&\left.\frac{2B^{\theta}f_R^\theta}{B^3r^2}\right]-\frac{C'}{BC}\left[
\frac{f_R^\theta}{B^2r^2}\left(\frac{B^\theta}{B}-\frac{C^\theta}{C}\right)
-\frac{\dot{f_R}}{A^2}\left(\frac{\dot{B}}{B}+\frac{\dot{C}}{C}\right)\right.\\\label{A2}
&-&\left.\frac{f'_R}{B^2}\left(\frac{B'}{B}+\frac{C'}{C}\right)+\frac{f''_R}{B^2}\right],
\\\nonumber
D_2&=&B\left[\frac{(-1)}{A^2B^2r^2}\left(\dot{f^\theta_R}-\frac{A^\theta}{A}\dot{f_R}
-\frac{\dot{B}}{B}f^\theta_R\right)\right]_{,0}+B^2
\left[\frac{1}{B^4r^2}
\left(f'^\theta_R-\frac{B^\theta}{B}f'_R\right.\right.\\\nonumber
&-&\left.\left.\frac{f^\theta_R}{r}
-\frac{B'}{B}f^\theta_R\right)\right]_{,1}-\frac{1}{A^2r^2}
\left(\dot{f^\theta_R}
-\frac{A^\theta}{A}\dot{f_R}-\frac{\dot{B}}{B}f^\theta_R\right)\left(
\frac{3\dot{B}}{B}+\frac{\dot{C}}{C}\right)\\\nonumber
&+&\frac{1}{B^2r^2}
\left(f'^\theta_R-\frac{B^\theta}{B}f'_R-\frac{f^\theta_R}{r}
-\frac{B'}{B}f^\theta_R\right)\left(\frac{A'}{A}+\frac{3}{r}
+\frac{4B'}{B}+\frac{C'}{C}\right)\\\nonumber
&+&\frac{A^\theta}{Ar^2}\left[
\frac{f^{\theta\theta}}{B^2r^2}+\frac{\ddot{f_R}}{A^2}
-\frac{\dot{f_R}}{A^2}\left(\frac{\dot{A}}{A}+\frac{\dot{B}}{B}\right)
+\frac{f'_R}{B^2}\left(\frac{1}{r}-\frac{B'}{B}-\frac{A'}{A}\right)\right.
\\\nonumber
&-&\left.\frac{f_R^\theta}{B^2r^2}\left(\frac{A^\theta}{A}+\frac{B^\theta}{B}\right)
\right]+\frac{B^\theta}{Br^2}\left[\frac{f_R^{\theta\theta}}{B^2r^2}
-\frac{f''_R}{B^2}-\frac{f'_R}{B^2r}+\frac{\dot{f_R}}{A^2}\left(
\frac{2\dot{C}}{C}\right)\right.\\\nonumber
&-&\left.\frac{f^\theta_R}{B^2r^2}\left(\frac{2B^\theta}{B}
\right)\right]-\frac{1}{r^2}\left[\frac{f-Rf_R}{2}-\frac{\ddot{f_R}}{A^2}
+\frac{f''_R}{B^2}+\frac{\dot{f_R}}{A^2}\left(\frac{\dot{A}}{A}
-\frac{\dot{B}}{B}-\frac{\dot{C}}{C}\right)\right.\\\nonumber
&+&\left.\frac{f'_R}{B^2}
\left(\frac{A'}{A}+\frac{C'}{C}-\frac{B'}{B}\right)
+\frac{f_R^{\theta}}{B^2r^2}\left(\frac{A^\theta}
{A}+\frac{C^\theta}{C}-\frac{B^\theta}{B}\right)\right]_{,2}
+\frac{C^\theta}{Cr^2}\left[\frac{f_R^{\theta\theta}}{B^2r^2}\right.\\\label{A3}
&+&\left.\frac{\dot{f_R}}{A^2}\left(\frac{\dot{C}}{C}
-\frac{\dot{B}}{B}\right)-\frac{f'_R}{B^2}
\left(\frac{B'}{B}+\frac{C'}{C}+\frac{1}{r}\right)
-\frac{f_R^{\theta}}{B^2r^2}\left(\frac{B^\theta}
{B}+\frac{C^\theta}{C}\right)\right].
\end{eqnarray}
The static portions of Eqs.(\ref{33}) and (\ref{34}) are computed as
\begin{eqnarray}\nonumber
D_{1S}&=&\frac{-2{\epsilon}}{B_0}\left[-\frac{R_0^2}{4}
+\frac{R_0^{\theta\theta}}{B_0^2r^2}+\frac{R'_0}{B_0^2}
\left(\frac{A'_0}{A_0}-\frac{B'_0}{B_0}-\frac{1}{r}+\frac{C'_0}{C_0}\right)
+\frac{R_0^\theta}{B_0^2r^2}\right.\\\nonumber
&\times&\left.\left(\frac{A_0^\theta}{A_0}
-3\frac{B_0^\theta}{B_0}+\frac{C_0^\theta}{C_0}\right)\right]_{,1}
+\frac{2{\epsilon}}{B_0^3r^2}\left(\frac{A_0^\theta}{A_0}
+\frac{4B_0^\theta}{B_0}\right)\left(R_0^\theta-R_0\frac{
B_0^\theta}{B_0}\right.\\\nonumber
&-&\left.\frac{R_0^\theta}{r}-R_0^\theta\frac{B'_0}{B_0}\right)+2{\epsilon}B_0
\left[R'^\theta_0-\frac{B_0^\theta}{B_0}R_0
-\frac{R_0^\theta}{r}-R_0^\theta\frac{B'_0}{B_0}\right]_{,2}
+\frac{2{\epsilon}A'_0}{A_0B^3_0}\\\nonumber
&\times&\left[R''_0-\frac{R_0^{\theta\theta}}{r^2}
+\frac{R_0^\theta}{r^2}\left(\frac{B_0^\theta}{B_0}-\frac{A^\theta_0}{A_0}\right)
+R'_0\left(\frac{2}{r}-\frac{B'_0}{B_0}-\frac{A'_0}{A_0}\right)\right]
+\frac{2{\epsilon}}{B_0^2}\\\nonumber
&\times&\left(\frac{1}{r}+\frac{B'_0}{B_0}\right)
\left[R''_0-\frac{R_0^{\theta\theta}}{r^2}-\frac{R'_0}{r}+\frac{R_0^\theta}{r^2}
\left(\frac{2B_0^\theta}{B_0}\right)\right]-\frac{2{\epsilon}C'_0}{C_0B^3_0}
\left[\frac{R_0^\theta}{r^2}\left(\frac{B_0^\theta}{B_0}\right.\right.\\\label{A4}
&-&\left.\left.\frac{C_0^\theta}{C_0}\right)-R'_0\left(\frac{B_0'}{B_0}
+\frac{C_0'}{C_0}\right)+R''_0\right], \\\nonumber
D_{2S}&=&B_0^2\left[\frac{2{\epsilon}}{B_0^4r^2}
\left\{R'^\theta_0-\frac{B_0^\theta}{B_0}R_0
-\frac{R_0^\theta}{r}-R_0^\theta\frac{B'_0}{B_0}\right\}\right]_{,1}
+\frac{2{\epsilon}}{B_0^2r^2}\left(R'^\theta_0-\frac{B_0^\theta}{B_0}R_0\right.\\\nonumber
&-&\left.\frac{R_0^\theta}{r}-R_0^\theta\frac{B'_0}{B_0}\right)
\left(\frac{A'_0}{A_0}+\frac{4B'_0}{B_0}+\frac{3}{r}+\frac{C'_0}{C_0}\right)
+\frac{2{\epsilon}A_0^\theta}{rA_0B_0^2}\left[\frac{R_0^{\theta\theta}}{r^2}
+R'_0\left(\frac{1}{r}\right.\right.\\\nonumber
&-&\left.\left.\frac{B'_0}{B_0}-\frac{A'_0}{A_0}\right)
-\frac{R_0^\theta}{r^2}\left(\frac{B_0^\theta}{B_0}
+\frac{A_0^\theta}{A_0}\right)\right]+\frac{2{\epsilon}B_0^\theta}
{B_0^3r^2}\left[\frac{R_0^{\theta\theta}}{r^2}-R''_0-\frac{R'_0}{r}
-\frac{R_0^\theta}{r^2}\right.\\\nonumber
&\times&\left.\left(\frac{2B_0^\theta}{B_0}\right)\right]-\frac{2{\epsilon}}{B_0^2r^2}
\left[-\frac{B_0^2R_0^2}{4}+R''_0+R'_0
\left(\frac{C'_0}{C_0}-\frac{B'_0}{B_0}+\frac{A'_0}{A_0}\right)
+\frac{R_0^\theta}{r^2}\right.\\\nonumber
&\times&\left.\left(\frac{A_0^\theta}{A_0}-\frac{B_0^\theta}{B_0}
+\frac{C_0^\theta}{C_0}\right)\right]_{,2}+\frac{2{\epsilon}C_0^\theta}{C_0B_0^2r^2}
\left[\frac{R_0^{\theta\theta}}{r^2}-R'_0
\left(\frac{1}{r}+\frac{B'_0}{B_0}+\frac{C'_0}{C_0}\right)\right.\\\label{A5}
&-&\left.\frac{R_0^\theta}{r^2}\left(\frac{B_0^\theta}{B_0}
+\frac{C_0^\theta}{C_0}\right)\right].
\end{eqnarray}
The perturbed portions of Eqs.(\ref{35})-(\ref{37}) are
\begin{eqnarray}\nonumber
D_3&=&-{\epsilon}eR_0+\frac{2{\epsilon}}{B_0^2}\left(e''
-\frac{2bR''_0}{B_0}\right)+\frac{2{\epsilon}}{B_0^2r^2}
\left(e^{\theta\theta}-\frac{2bR^{\theta\theta}_0}
{B_0}\right)+\frac{2{\epsilon}}{B_0^2}\left(e'\right.\\\nonumber
&-&\left.\frac{2bR'_0}{B_0}\right)\left(\frac{C'_0}{C_0}
-2\frac{B'_0}{B_0}+\frac{1}{r}\right)+\frac{2{\epsilon}
R'_0}{B_0^2}\left\{\left(\frac{c}{C_0}\right)'-2\left(
\frac{b}{B_0}\right)'\right\}+\frac{2{\epsilon}}{B_0^2r^2}\\\nonumber
&\times&\left(e^{\theta}-\frac{2bR^{\theta}_0}{B_0}\right)
\left(\frac{C_0^\theta}{C_0}-2\frac{B_0^\theta}{B_0}\right)
+\frac{2{\epsilon}R_0^\theta}{B^2r^2}\left\{\left(\frac{c}{C_0}
\right)^\theta-2\left(\frac{b}{B_0}\right)^\theta\right\}
+\frac{2{\epsilon}}{B_0^2}\\\nonumber
&\times&\left(e'-\frac{eA'_0}{A_0}
-b\frac{R'_0}{B_0}\right)\left(3\frac{A'_0}{A_0}-2\frac{B'_0}{B_0}
+\frac{1}{r}+\frac{C'_0}{C_0}\right)-\frac{2{\epsilon}}{B_0^2r^2}
\left(e^\theta-\frac{eA^\theta_0}{A_0}\right.\\\nonumber
&-&\left.b\frac{R^\theta_0}{B_0}\right)\left(3\frac{A^\theta_0}
{A_0}-2\frac{B^\theta_0}{B_0}+\frac{C^\theta_0}{C_0}\right)
+A_0^2\left[-\frac{2{\epsilon}}{A_0^2B_0^2}
\left(e'-\frac{eA'_0}{A_0}-b\frac{R'_0}{B_0}\right)\right]_{,1}\\\nonumber
&+&A_0^2\left[\frac{-1}{A_0^2B_0^2r^2}\left(e^\theta
-\frac{eA^\theta_0}{A_0}-b\frac{R^\theta_0}{B_0}\right)
\right]_{,2}+\frac{2{\epsilon}b}{B_0^3}
\left[R''_0+\frac{R^{\theta\theta}_0}{r^2}-2R_0^\theta
\left(2\frac{A'_0}{A_0}\right.\right.\\\nonumber
&+&\left.\left.2\frac{B'_0}{B_0}+\frac{3}{r}\right)
-\frac{A_0^{\theta}R_0^\theta}{A_0r^2}\right]
+\frac{2c{\epsilon}}{C_0B_0^2}\left[R'_0\left(\frac{C'_0}
{C_0}+\frac{2}{r}-\frac{A'_0}{A_0}\right)
+\frac{R_0^\theta}{r^2}\right.\\\label{A6}
&\times&\left.\left(\frac{C_0^\theta}{C_0}-\frac{A^\theta_0}
{A_0}\right)\right], \\\nonumber
P_1&=&-T\left[-{\epsilon}eR_0+2\frac{{\epsilon}R'_0}{B_0^2r^2}
\left(e^\theta-\frac{2bR_0^\theta}{B_0}\right)+\frac{2{
\epsilon}}{B_0^2}\left(e'-\frac{2bR_0'}{B_0}\right)
\left(\frac{A'_0}{A_0}-\frac{B'_0}{B_0}\right.\right.\\\nonumber
&-&\left.\left.\frac{1}{r}+\frac{C'_0}{C_0}\right)
+\frac{2{\epsilon}}{B_0^2}
\left\{\left(\frac{a}{A_0}\right)'-\left(\frac{b}{B_0}\right)'
+\left(\frac{c}{C_0}\right)'\right\} +\frac{2{\epsilon}}{B_0^2r^2}
\left(e^\theta-\frac{2bR_0^\theta}{B_0}\right)\right.\\\nonumber
&\times&\left.\left(\frac{A_0^\theta}{A_0}
-3\frac{B_0^\theta}{B_0}+\frac{C_0^\theta}{C_0}\right)
+\frac{2{\epsilon}R_0^\theta}{B_0^2r^2}\left\{\left(\frac{a}{A_0}
\right)^\theta-\left(\frac{b}{B_0}\right)^\theta
+\left(\frac{c}{C_0}\right)^\theta\right\}\right]\\\nonumber
&+&\frac{2{\epsilon}e\ddot{T}}{A_0^2B_0}+\frac{T}{r^2B_0^2}
\left[\left\{\left(\frac{a}{A_0}\right)^\theta+4\left(
\frac{b}{B_0}\right)^\theta\right\}\left(2{\epsilon}
R'^\theta_0-2{\epsilon}\frac{R_0^\theta}{r}
-2{\epsilon}R'_0\right.\right.\\\nonumber
&\times&\left.\left.\frac{B_0^\theta}{B_0}-2{\epsilon}R_0^\theta
\frac{B'_0}{B_0}\right)+\left(\frac{A_0^\theta}{A_0}
+\frac{4B_0^\theta}{B_0}\right)2{\epsilon}\left\{ -e'
\frac{B_0^\theta}{B_0}-R'_0\left(\frac{b}{B_0}\right)^\theta
-\frac{e^\theta}{r}\right.\right.\\\nonumber
&-&\left.\left.e^\theta\frac{4B_0'}{B_0}-R_0^\theta
\left(\frac{b}{B_0}\right)'\right\}\right]- \frac{2T}{B_0^3r^2}
\left[\left(\frac{A_0^\theta}{A_0}+\frac{4B_0^\theta}
{B_0}\right)2{\epsilon}\left(R'^\theta_0
-\frac{R_0^\theta}{r}\right.\right.\\\nonumber
&-&\left.\left.R'_0\frac{B_0^\theta}{B_0}-R_0^\theta
\frac{B_0'}{B_0}\right)\right]+\ddot{T}\left[\frac{
-2{\epsilon}}{A_0^2}\left\{e'-\frac{A'_0}{A_0}e
-\frac{b}{B_0}R'_0\right\}\right]-2{\epsilon}TB_0^2\\\nonumber
&\times&\left[\frac{1}{B_0^2r^2}\left\{e'\frac{B_0^\theta}
{B_0}+R'_0\left(\frac{b}{B_0}\right)^\theta
+\frac{e^\theta}{r}+e^\theta\frac{B'_0}{B_0}
+R_0^\theta\left(\frac{b}{B_0}\right)'\right\}\right.\\\nonumber
&+&\left.\frac{4b}{B_0^5r^2}
\left\{R'^2_0-\frac{R_0^\theta}{r}-R'_0\frac{B_0^\theta}{B_0}
-R_0\frac{B'_0}{B_0}\right\}\right]_{,2}+\frac{2T{\epsilon}}{B_0}
\left(\frac{a}{A_0}\right)'\left[R''_0-\frac{R_0^{\theta\theta}}
{r^2}\right.\\\nonumber&+&\left.\frac{R_0^{\theta}}{r^2}
\left(\frac{B_0^\theta}{B_0}-\frac{A_0^\theta}{A_0}\right)
+R'_0\left(\frac{2}{r}-\frac{B_0'}{B_0}-\frac{A_0'}{A_0}\right)\right]
+\frac{2{\epsilon}TA'_0}{A_0B_0^2}\left[e''-\frac{2bR''_0}{B_0}\right.\\\nonumber
&-&\left.\frac{1}{r^2}\left(e^{\theta\theta}-\frac{2bR^{\theta\theta}_0}{B_0}\right)
+\frac{R_0^\theta}{r^2}\left\{\left(\frac{b}{B_0}\right)^\theta
-\left(\frac{a}{A_0}\right)^\theta\right\}+\frac{1}{r^2}
\left(e^{\theta}-\frac{2bR^\theta_0}{B_0}\right)\right.\\\nonumber
&\times&\left.\left(\frac{B_0^\theta}{B_0}-\frac{A_0^\theta}{A_0}\right)
+R'_0\left\{\frac{2}{r}-\left(\frac{b}{B_0}\right)'
-\left(\frac{a}{A_0}\right)'\right\}+\left(e'
-\frac{2bR_0'}{B_0}\right)\left(\frac{2}{r}\right.\right.\\\nonumber
&-&\left.\left.\frac{B'_0}{B_0}
-\frac{A'_0}{A_0}\right)\right]+\frac{2{\epsilon}}{B_0^2}
\left(\frac{1}{r}+\frac{B'_0}{B_0}\right)\left[e''-\frac{2bR''_0}{B_0}
-\frac{1}{r^2}\left(e^{\theta\theta}-\frac{2bR^{\theta\theta}_0}
{B_0}\right)\right.\\\nonumber
&-&\left.\frac{1}{r}\left(e'-\frac{2bR'_0}{B_0}\right)
-\frac{2R_0^\theta}{r^2}\left(\frac{b}{B_0}\right)^\theta
+\frac{2B_0^\theta}{B_0r^2}\left(e^\theta
-\frac{2bR_0^\theta}{B_0}\right)\right]+\frac{2{\epsilon}T}{B_0^2}\\\nonumber
&\times&\left(\frac{b}{B_0}\right)'\left[R''_0-\frac{R_0^{\theta\theta}}{r^2}
-\frac{R'_0}{r^2}+\frac{R_0^\theta}{r^2}\left(2B_0^\theta\right)\right]
-\frac{2{\epsilon}TC'_0}{B_0^2C_0}\left[e''-\frac{2bR''_0}{B_0}\right.\\\nonumber
&+&\left.\frac{R_0^\theta}{r^2}\left\{\left(\frac{b}{B_0}\right)^\theta
-\left(\frac{c}{C_0}\right)^\theta\right\}+\left(e^\theta
-\frac{2bR_0^\theta}{B_0}\right)\left(\frac{B_0^\theta}{B_0}
-\frac{C_0^\theta}{C_0}\right)-R'_0\right.\\\nonumber
&\times&\left.\left\{\left(\frac{b}{B_0}\right)'+\left(\frac{c}{C_0}
\right)'\right\}-\left(e'-\frac{2bR_0'}{B_0}\right)
\left(\frac{B'_0}{B_0}+\frac{C'_0}{C_0}\right)\right]
-\frac{2{\epsilon}T}{B_0^2}\left(\frac{c}{C_0}\right)'\\\label{A7}
&\times&\left[R''_0+\frac{R_0^\theta}{r^2}\left(\frac{B_0^\theta}{B_0}
-\frac{C_0^\theta}{C_0}\right)-R'_0\left(\frac{B'_0}{B_0}
+\frac{C'_0}{C_0}\right)\right], \\\nonumber
D_5&=&TB_0^2\left[\frac{2{\epsilon}}{B_0^4r^4}\left\{e'^\theta
-\frac{e^\theta}{r}-R'_0\left(\frac{b}{B_0}\right)^\theta
-e'\frac{B_0^\theta}{B_0}-R_0^\theta\left(\frac{b}{B_0}\right)'
\right\}\right.\\\nonumber
&-&\left.\frac{8{\epsilon}b}{B_0^5r^4}\left(R'^\theta_0
-\frac{R_0^\theta}{r}-R'_0\frac{B_0^\theta}{B_0}-R_0^\theta
\frac{B'_0}{B_0}\right)\right]_{,1}+4Tb{\epsilon}B_0
\left[\frac{1}{B_0^4r^2}\right.\\\nonumber
&\times&\left.\left(R'^\theta_0
-\frac{R_0^\theta}{r}-R'_0\frac{B_0^\theta}{B_0}-R_0^\theta
\frac{B'_0}{B_0}\right)\right]_{,1}+\frac{2{\epsilon}TB_0^\theta}{B_0^3r^2}
\left[\frac{1}{r^2}\left(e^{\theta\theta}\frac{2bR_0^{\theta\theta}}
{B_0}\right)\right.\\\nonumber
&-&\left.e''+\frac{2bR''_0}{B_0}-\frac{1}{r}
\left(e'-\frac{2bR'_0}{B_0}\right)-\frac{2R_0^\theta}{r^2}
\left(\frac{b}{B_0}\right)^{\theta}-\frac{2}{r^2}
\left(e^\theta-\frac{2bR_0^\theta}{B_0}\right)\right.\\\nonumber
&\times&\left.\frac{B_0^\theta}{B_0}\right]
+\frac{2{\epsilon}T}{B_0^2r^2}\left(\frac{b}{B_0}\right)^\theta
\left[\frac{R_0^{\theta\theta}}{r^2}-R''_0-\frac{R'_0}{r}
-\frac{2R_0^\theta}{r^2}\left(\frac{B_0^\theta}{B_0}\right)\right]
-\frac{2{\epsilon}T}{B_0^2r^2}\\\nonumber
&\times&\left(\frac{a}{A_0}\right)^\theta
\left[\frac{R_0^{\theta\theta}}{r^2}+R'_0\left(\frac{1}{r}-\frac{B'_0}{B_0}
-\frac{A'_0}{A_0}\right)-\frac{R_0^\theta}{r^2}
\left(\frac{A_0^\theta}{A_0}+\frac{B_0^\theta}{B_0}\right)\right]\\\nonumber
&+&\frac{2{\epsilon}TA_0^\theta}{A_0B_0^2r^2}\left[\frac{1}{r^2}
\left(e^{\theta\theta}-\frac{2bR_0^{\theta\theta}}{B_0}\right)
-R'_0\left\{\left(\frac{b}{B_0}\right)'+\left(\frac{a}{A_0}\right)'
\right\}\right.\\\nonumber
&+&\left.\left(e'-\frac{2bR'_0}{B_0}\right)\left(\frac{1}{r}-\frac{B'_0}{B_0}
-\frac{A'_0}{A_0}\right)-\frac{R_0^\theta}{r^2}
\left\{\left(\frac{b}{B_0}\right)^\theta+\left(\frac{a}{A_0}\right)^\theta
\right\}\right.\\\nonumber
&-&\left.\frac{1}{r^2}\left(e^\theta-\frac{2bR\theta_0}{B_0}\right)
\left(\frac{A_0^\theta}{A_0}+\frac{B_0^\theta}{B_0}\right)\right]
+\frac{2{\epsilon}eA'_0\ddot{T}}{A_0^3r^2}+\frac{2{\epsilon}}{B_0^2r^2}
\left(R'^\theta_0\right.\\\nonumber
&-&\left.\frac{R_0^\theta}{r}-R'_0\frac{B_0^\theta}{B_0}-R_0^\theta
\frac{B'_0}{B_0}\right)\left\{\left(4\frac{b}{B_0}\right)'+\left(\frac{a}{A_0}\right)'
+\left(\frac{c}{C_0}\right)'\right\}\\\nonumber
&+&\frac{2T{\epsilon}}{B_0^2r^2} \left\{e'^\theta
-\frac{e^\theta}{r}-R'_0\left(\frac{b}{B_0}\right)^\theta
-e'\frac{B_0^\theta}{B_0}-R_0^\theta\left(\frac{b}{B_0}\right)'
\right\}\left(\frac{A'_0}{A_0}\right.\\\nonumber
&+&\left.\frac{3}{r}+4\frac{B'_0}{B_0}+\frac{C'_0}{C_0}
\right)-\frac{4{\epsilon}Tb}{B_0^3r^2}\left(R'^\theta_0
-\frac{R_0^\theta}{r}-R'_0\frac{B_0^\theta}{B_0}-R_0^\theta
\frac{B'_0}{B_0}\right)\\\nonumber
&\times&\left(\frac{A'_0}{A_0}+\frac{3}{r}
+4\frac{B'_0}{B_0}+\frac{C'_0}{C_0}\right)-\frac{2{\epsilon}T}{r^2}
\left[\frac{1}{B_0^2}\left\{-\frac{eR_0B_0^2}{2}+e''\right.\right.\\\nonumber
&-&\left.\left.\frac{2bR''_0}{B_0}
+R'_0\left\{\left(\frac{a}{A_0}\right)'-\left(\frac{b}{B_0}\right)'
+\left(\frac{c}{C_0}\right)'\right\}+\left(e'-\frac{2bR'_0}{B_0}\right)
\right.\right.\\\nonumber &\times&\left.\left.
\left(\frac{A'_0}{A_0}-\frac{B'_0}{B_0}+\frac{C'_0}{C_0}\right)
+\frac{R_0^\theta}{r^2}\left\{\left(\frac{a}{A_0}\right)^\theta
-\left(\frac{b}{B_0}\right)^\theta+\left(\frac{c}{C_0}
\right)^\theta\right\}\right.\right.\\\nonumber
&+&\left.\left.\frac{1}{r^2}\left(\frac{A^\theta_0}{A_0}
-\frac{B^\theta_0}{B_0}+\frac{C^\theta_0}{C_0}\right)\right\}\right]_{,2}
+\frac{2{\epsilon}C_0^\theta}{C_0B_0^2r^2}\left[\frac{1}{r^2}\left(e^{\theta\theta}
-\frac{2bR_0^{\theta\theta}}{B_0}\right)\right.\\\nonumber
&-&\left.R'_0\left\{\left(\frac{b}{B_0}\right)'
+\left(\frac{c}{C_0}\right)'\right\}+\left(e'-\frac{2bR_0'}
{B_0}\right)\left(\frac{B'_0}{B_0}+\frac{1}{r}+\frac{C'_0}{C_0}\right)
\right.\\\nonumber
&-&\left.\frac{R_0^\theta}{r^2}\left\{\left(\frac{b}{B_0}\right)^\theta
+\left(\frac{c}{C_0}\right)^\theta\right\}+\frac{1}{r^2}
\left(e^{\theta} -\frac{2bR_0^{\theta}}{B_0}\right)\left(
\frac{B_0^\theta}{B_0}+\frac{C_0^\theta}{C_0}\right)\right]\\\label{A8}
&+&\frac{2{\epsilon}T}{B_0^2r^2}\left[\frac{R_0^{\theta\theta}}{r^2}
-R'_0\left(\frac{B'_0}{B_0}+\frac{1}{r}+\frac{C'_0}{C_0}\right)
-\frac{R_0^\theta}{r^2}\left(B_0^\theta+\frac{C_0^\theta}{C_0}\right)\right].
\end{eqnarray}
The coefficient $\delta^2$ of the differential equation (\ref{39})
is given by
\begin{eqnarray}\nonumber
\delta^2&=&\frac{-A_0^2}{2}\left[e
-\frac{2}{B_0^2}\left\{\frac{C_0'A_0'}{C_0A_0}\left(
\frac{a'}{A_0'}-\frac{a}{A_0}+\frac{c'}{C_0'}-\frac{c}{C_0}\right)
+\left(\frac{a''}{A_0''}-\frac{a}{A_0}\right)\frac{A_0''}
{A_0}\right.\right.\\\nonumber
&+&\left.\left(\frac{b''}{B_0''}-\frac{b}{B_0}\right)\frac{B_0''}{B_0}
-\frac{1}{r}\left(\frac{a}{A_0}-\frac{c}{C_0}-\frac{b}{B_0}\right)'
+\left(\frac{\bar{c}''}{C_0''}-\frac{\bar{c}}{C_0}\right)\frac{C_0''}
{C_0}-2\left(\frac{b}{B_0}\right)'\right.\\\nonumber
&\times&\left.\frac{B_0'}{B_0}+\left(\frac{b}{B_0}\right)^{\theta}
\frac{B^\theta_0}{B_0}\frac{2}{r^2}+\left(\frac{c^{\theta\theta}}
{C^{\theta\theta}_0}-\frac{c}{C_0}\right)\frac{C^{\theta\theta}_0}
{C_0}+\left(\frac{b^{\theta\theta}}{B^{\theta\theta}_0}-\frac{b}{B_0}
\right)\frac{B^{\theta\theta}_0}{B_0}+\left(\frac{a^{\theta\theta}}
{A^{\theta\theta}_0}\right.\right.\\\nonumber
&-&\left.\left.\left.\frac{a}{A_0}
\right)\frac{A^{\theta\theta}_0}{A_0}+\frac{C^{\theta}_0A^{\theta}_0}
{C_0A_0}\left(\frac{a^\theta}{A^{\theta}_0}-\frac{a}{A_0}
+\frac{c^\theta}{C^{\theta}_0}-\frac{c}{C_0}\right)\right\}
+2\frac{R_0b}{B_0}\right]\left(\frac{b}{B_0}
-\frac{\bar{c}}{C_0}\right)^{-1}.\\\label{A9}
\end{eqnarray}

\vspace{0.25cm}

\acknowledgments


\nocite{*}

\begin{thebibliography}{}

\bibitem[Allen et al.(2004)]{t4}Allen, S.W., Schmidt, R.W., Ebeling, H., Fabian, A.C. and
Speybroeck, L.V.: Mon. Not. R. Astron. Soc. \textbf{353}, 457 (2004)

\bibitem[Amendola et al.(2007)]{t26c} Amendola, L., Gannouji, R., Polarski, D. and Tsujikawa, S.: Phys.
Rev. D \textbf{75}, 083504 (2007)

\bibitem[Bennett et al.(2003)]{t3}Bennett, C.L. et al.: Astrophys. J. Suppl. Ser. \textbf{148},
1 (2003)

\bibitem[Bergliaffa and Nunes(2011)]{t19} Bergliaffa, S.E.P. and Nunes, Y.E.C.O.: Phys. Rev. D
\textbf{84}, 084006 (2011)

\bibitem[Capozziello(2002)]{t6}Capozziello, S.: Int. J. Mod. Phys. D \textbf{11}, 483 (2002)

\bibitem[Capozziello and Laurentis(2011)]{t27} Capozziello, S. and Laurentis, M.D.: Phys. Rep.
\textbf{509}, 167 (2011)

\bibitem[Capozziello et al.(2011)]{t124} Capozziello, S., De Laurentis, M., Odintsov, S.D. and Stabile, A.:
Phys. Rev. D \textbf{83}, 064004 (2011)

\bibitem[Capozziello et al.(2012)]{t125} Capozziello, S., De Laurentis, M., De Martino, I., Formisano, M. and
Odintsov, S.D.: Phys. Rev. D \textbf{85}, 044022 (2012)

\bibitem[Cembranos et al.(2012)]{t32} Cembranos, J.A.R., Dombriz, A.D.L.C. and
N\'{u}\~{n}ez, B.M.: J. Cosmol. Astropart. Phys. \textbf{04}, 021
(2012)

\bibitem[Chakraborty et al.(2005)]{t12}Chakraborty, S., Chakraborty, S. and Debnath, U.:
Int. J. Mod. Phys. D \textbf{14}, 1707 (2005)

\bibitem[Chandrasekhar(1964)]{t21} Chandrasekhar, S.: Astrophys. J. \textbf{140}, 417 (1964)

\bibitem[Chan(2000)]{t23c} Chan, R.: Mon. Not. R. Astron. Soc. \textbf{316}(2000)588

\bibitem[Chan et al.(1993)]{t23a}Chan, R., Herrera, L. and Santos, N.O.: Mon. Not. R. Astron. Soc.
\textbf{265}, 533 (1993)

\bibitem[Chan et al.(1994)]{t23b}Chan, R., Herrera, L. and Santos, N.O.: Mon. Not. R. Astron. Soc.
\textbf{267}, 637 (1994)

\bibitem[Cognola et al.(2005)]{t17}Cognola, G., Elizalde, E., Nojiri, S., Odintsov, S.D. and
Zerbini, S.: J. Cosmol. Astropart. Phys. \textbf{02}, 010 (2005)

\bibitem[Cognola et al.(2008)]{t127}Cognola, G., Elizalde, E., Nojiri, S., Odintsov, S.D., Sebastiani,
L. and Zerbini, S.: Phys. Rev. D \textbf{77}, 046009 (2008)

\bibitem[Copeland et al.(2006)]{t26a}Copeland, E.J., Sami, M. and Tsujikawa, S.: Int. J. Mod. Phys. D
\textbf{15}, 1753 (2006)

\bibitem[Cruz-Dombriz, A.D.L. and S\'{a}ez-G\'{o}mez(2012)]{t131}Cruz-Dombriz, A.D.L. and S\'{a}aez-G\'{o}mez,
D.: Entropy \textbf{14}, 1717 (2012)

\bibitem[Cruz-Dombriz et al.(2009)]{t18}Cruz-Dombriz, A.D.L., Dobado, A. and Maroto, A.L.: Phys. Rev. D
\textbf{80}, 124011 (2009)

\bibitem[Farinelli et al.(2014)]{t126}Farinelli, R, De Laurentis, M., Capozziello, S. and Odintsov, S.D.:
arXiv:1311.2744 (2014)

\bibitem[Felice and Tsujikawa(2010)]{t6a}Felice, A.D. and Tsujikawa, S.: Living Rev. Relativity
\textbf{13}, 3 (2010)

\bibitem[Garattini(2009)]{t14a}Garattini R.: J. Phys.: Conf. Ser., \textbf{174}, 012066 (2009)

\bibitem[Harrison et al.(1965)]{t34} Harrison, B.K., Thorne, K.S., Wakano, M. and Wheeler,
J.A.: \textit{Gravitation Theory and Gravitational Collapse}
(University of Chicago Press, 1965)

\bibitem[Herrera and Santos(2010)]{t8}Herrera, L. and Santos, N.O.: Phys. Rep.
\textbf{286}, 53 (1997)

\bibitem[Herrera and Santos(1995)]{t10}Herrera, L. and Santos, N.O.: Astrophys. J. \textbf{438}, 308 (1995)

\bibitem[Herrera et al.(1989)]{t22} Herrera, L., Santos, N.O. and Le Denmat, G.: Mon. Not. R. Astron. Soc.
\textbf{237}, 257 (1989)

\bibitem[Herrera et al.(2013)]{t28} Herrera, L., Di Prisco, A.: J. Ospino and J. Ib\'{a}\~{n}ez, Phys.
Rev. D \textbf{87}, 024014 (2013)

\bibitem[Herrera et al.(2014)]{t28her} Herrera, L., Di Prisco, A., Ib¶a~nez, J., Ospino, J.: Phys.
Rev. D \textbf{89}, 084034 (2014)

\bibitem[Huang(2014)]{tpoly} Huang, Q.G.: J. Cosmol. Astropart. Phys., \textbf{02}, 035 (2014)

\bibitem[Jain and Taylor(2003)]{t2}Jain, B. and Taylor, A.: \prl~\textbf{91},
141302 (2003)

\bibitem[Mak and Harko(2003)]{t8a}Mak, M.K. and Harko, T.: R. Soc. London A
\textbf{459}, 393 (2003)

\bibitem[Matarrese and Terranova(1996)]{t24} Matarrese, S. and Terranova, D.: Mon. Not. R. Astron. Soc.
\textbf{283}, 400 (1996)

\bibitem[Misner and Sharp(1964)]{t14}Misner, C.W. and Sharp, D.:
Phys. Rev. \textbf{136}, B571 (1964)

\bibitem[Nojiri and Odintsov(2007)]{t26b} Nojiri, S. and Odintsov, S.D.: Int. J. Geom. Meth. Mod. Phys.
\textbf{4}, 115 (2007)

\bibitem[Nojiri and Odintsov(2011)]{t128}Nojiri, S. and Odintsov, S.D.: Phys. Rep. \textbf{505}, 59 (2011)

\bibitem[Nojiri and Odintsov(2013)]{t130}Nojiri, S. and Odintsov, S.D.: Class. Quantum Grav. \textbf{30},
125003 (2013)

\bibitem[Oppenheimer and Snyder(1939)]{t13}Oppenheimer, J.R. and Snyder, H.:
Phys. Rev. \textbf{56}, 455 (1939)

\bibitem[Perlmutter et al.(1998)]{t1a}Perlmutter, S. et al.: Nature \textbf{391},
51 (1998)

\bibitem[ Riess et al.(1998)]{t1} Riess, A.G. et al.: Astrophys. J. \textbf{116},
1009 (1998)

\bibitem[Roy and Tripathi(1971)]{t29}Roy, S.R. and Tripathi, V.N.: Gen. Realtiv. Gravit.
\textbf{2}, 121 (1971)

\bibitem[Sebastiani et al.(2013)]{t129}Sebastiani, L., Momeni, D., Myrzakulov, R. and Odintsov, S.D.:
Phys. Rev. D \textbf{88}, 104022 (2013)

\bibitem[Sharif and Bhatti(2012a)]{t15a}Sharif, M. and Bhatti, M.Z.: Gen. Relativ. Gravit.
\textbf{44}, 2811 (2012a).

\bibitem[Sharif and Bhatti(2012b)]{t15b}Sharif, M. and Bhatti, M.Z.: Mod. Phys. Lett. A \textbf{27}, 1250141 (2012b)

\bibitem[Sharif and Bhatti(2013a)]{t25a} Sharif, M. and Bhatti, M.Z.: J. Cosmol. Astropart. Phys.
\textbf{10}, 056 (2013a)

\bibitem[Sharif and Bhatti(2014)]{t25b} Sharif, M. and Bhatti, M.Z.: Phys. Lett. A \textbf{378}, 469 (2014)

\bibitem[Sharif and Bhatti(2013b)]{t26} Sharif, M. and Bhatti, M.Z.: J. Cosmol. Astropart. Phys.
\textbf{11}, 014 (2013b)

\bibitem[Sharif and Yousaf(2012a)]{t15c}Sharif, M. and Yousaf, Z.: Int. J. Mod. Phys. D \textbf{21},
1250095 (2012a)

\bibitem[Sharif and Yousaf(2012b)]{t15d} Sharif, M. and Yousaf, Z.: Can. J. Phys. \textbf{90}, 865 (2012b)

\bibitem[Sharif and Yousaf(2013)]{t20} Sharif, M. and Yousaf, Z.: Mon. Not. R. Astron. Soc.
\textbf{432}, 264 (2013)

\bibitem[Sharif and Yousaf(2013a)]{t25c} Sharif, M. and Yousaf, Z.: Phys. Rev. D \textbf{88}(2013a)024020.

\bibitem[Sharif and Yousaf(2014a)]{t30} Sharif, M. and Yousaf, Z.: Astrophys. Space Sci. \textbf{351}, 351 (2014a)

\bibitem[Sharif and Yousaf(2014b)]{tpoly1} Sharif, M. and Yousaf, Z.: Astrophys. Space Sci.
DOI 10.1007/s10509-014-1913-z (2014b)

\bibitem[Sotirou and Faraoni(2010)]{t33} Sotirou, T.P. and Faraoni, V.: Rev. Mod. Phys. \textbf{82}, 451 (2010)

\bibitem[Starobinsky(1980)]{t31} Starobinsky, A.A.: Phys. Lett. B \textbf{91}, 99 (1980)

\bibitem[Weber(1999)]{t10a}Weber, F.: \emph{Pulsars as Astrophysical Observatories for Nuclear and
Particle Physics}, IOP Publishing, Bristol, (1999)

\bibitem[Ziaie et al.(2011)]{t14b} Ziaie A. H., Atazadeh K. and Rasouli S.M.M.: Gen. Relativ.
Gravit., \textbf{43}, 2943 (2011)


\end{thebibliography}

\end{document}